\documentclass{elsart1p}
\usepackage{graphicx}

\usepackage{amssymb}
\usepackage{amsmath}
\renewcommand{\theenumi}{\arabic{enumi}}

\newcommand{\medio}[1]{\langle #1\rangle}
\newcommand{\aM}{\textsf{\itshape M\/}}
\newcommand{\aR}{\textsf{\itshape R\/}}
\newcommand{\aU}{\textsf{\itshape U\/}}
\newcommand{\am}{\textsf{\itshape m\/}}
\newcommand{\ar}{\textsf{\itshape r\/}}
\newcommand{\au}{\textsf{\itshape u\/}}
\newcommand{\ah}{\textsf{\itshape h\/}}
\newcommand{\aO}{\textsf{\itshape O\/}}
\newcommand{\aF}{\textsf{\itshape F\/}}
\newcommand{\hatam}{\hat{\am\,}\!}
\newcommand{\hatah}{\hat{\ah\,}\!}
\newcommand{\hataM}{\hat{\aM\,}\!}
\newcommand{\ee}{\mathrm{e}}
\newcommand{\dd}{\mathrm{d}}
\newcommand{\ii}{\mathrm{i}}

\begin{document}

\begin{frontmatter}

\journal{Nuclear Physics B}
\title{Tethered Monte Carlo: computing the effective potential without
critical slowing down} 
\author{L.A. Fernandez},
\author{V. Martin-Mayor},
\author{D. Yllanes\corauthref{cor}} 
\corauth[cor]{Corresponding author, email address: \texttt{yllanes@lattice.fis.ucm.es}}

\address{Departamento de F\'isica Te\'orica I, Facultad de CC. F\'isicas,\\
Universidad Complutense de Madrid, 28040 Madrid, Spain\\
and Instituto de Biocomputaci\'on y F\'isica\\ de Sistemas Complejos (BIFI), Zaragoza, Spain
}

\begin{abstract}
We present Tethered Monte Carlo, a simple,
general purpose method of computing the effective
potential of the order parameter (Helmholtz free energy).
This formalism is based on a new statistical ensemble,
closely related to the micromagnetic one, but with
an extended configuration space (through Creutz-like demons).
Canonical averages for arbitrary values of the external magnetic
field are computed without additional simulations.
The method is put to work in the two dimensional Ising 
model, where the existence of exact results 
enables us to perform high precision checks.
A rather peculiar feature of our implementation, 
which employs a local Metropolis algorithm, 
is the total absence, within errors, of critical
slowing down for magnetic observables. Indeed,
high accuracy results are presented for 
lattices as large as $L=1024$.
\end{abstract}

\begin{keyword}
Monte Carlo \sep Critical Phenomena \sep Phase transitions \sep Ising model

\PACS 75.40.Mg \sep 
      05.10.Ln \sep 
      05.50.+g      
\end{keyword}
\end{frontmatter}

\section{Introduction}
There exists a very profound relation between quantum field theory 
and statistical mechanics, through the theory of critical
phenomena~\cite{AMIT-VICTOR,ZINN-JUSTIN,PELISSETTO-VICARI}.
In the examination of these phenomena Monte Carlo simulations~\cite{MC} 
are an indispensable tool, because of their exceedingly broad
range of applicability. Indeed, Monte Carlo simulations not only succeed where an 
analytical treatment would be impossible or impractical, but they are many 
times the only numerical method capable of tackling the problem
at hand. 

There are some difficulties, though, among which we can mention
critical slowing down~\cite{ZINN-JUSTIN,SOKAL-TAU}. For traditional Monte Carlo formalisms
the correlation times (roughly, the number of intermediate steps so that two
measurements can be considered independent) grow as~$L^z$ at the critical
point (with $z\approx 2$). A great step towards the solution
of this problem was taken in the late 1980s, with the development of the first
cluster methods~\cite{CLUSTER}, capable of achieving $z<1$.
This prompted a large amount of work on more sophisticated cluster
algorithms, which continues today~\cite{CLUSTER-MODERNOS}. 

Unfortunately, cluster methods are highly specific and we do
not know how to efficiently implement them for physical problems as important
as lattice gauge theories~\cite{GAUGE}, with or without
dynamic fermions; structural glasses~\cite{GLASSES};
spin glasses~\cite{SPINGLASSES}; protein folding~\cite{PROTEIN}
and a long etcetera. Even in their favourite playground,
ferromagnetic systems, cluster methods lose most of their
power in the presence of a magnetic field. The simplest
example is the $D=2$ Ising model, whose exact solution without a magnetic
field is known since 1944~\cite{ONSAGER}, but
whose behaviour with a magnetic
perturbation is still an active research topic~\cite{ISING-H-ANALITICO,ISING-H-ANALITICO2}.
It is interesting to notice that the current numerical methods of choice 
rely on transfer matrix techniques~\cite{ISING-H-TM}
and not on Monte Carlo simulations. 

Here we present the Tethered Monte Carlo (TMC) method, 
a completely unspecific strategy, appliable to
many problems with or without an external field. 
The main goal of this approach is constructing the 
effective potential of the order parameter 
(perhaps more commonly named Helmholtz free energy
in a statistical mechanics context).
 
In order to define this Monte Carlo
method, we shall introduce a new statistical ensemble, 
where the magnetic field and the order parameter 
exchange their roles with respect
to the canonical ensemble.\footnote{In the 
canonical ensemble the magnetic field is an external parameter
and the order parameter an observable, 
while their tethered equivalents have the reversed roles.}
This new framework has some interesting features of its own.
For example, when working with a ferromagnetic system it provides
a very clean definition of the broken symmetry phase for a finite lattice.

In the ferromagnetic setting 
the tethered ensemble is related to the micromagnetical
one, where both the temperature $\beta$ and the magnetisation
density~$m$ are kept fixed. The difference is that here
we couple $m$ to a Gaussian `magnetostat'.
This yields a new variable $\hat m$ which, unlike $m$,
is continuous even for finite lattices. The magnetic field
is obtained from a fluctuation-dissipation formalism. We can then work 
at constant $\hat m$, where the real magnetisation is almost 
fixed, but has some leeway (hence the name \emph{tethered}).
We then combine the mean values at constant~$\hat m$ of the 
magnetic field to construct the effective potential $\varOmega_N(\hat m)$, 
whose exponential gives the canonical probability 
density function (pdf) of $\hat m$.
Using this pdf and the mean values of the different physical
observables as a function of $\hat m$ we can 
recover the canonical results. 

The tethered ensemble is not only a fancy theoretical construct,
but also the basis for a practical 
simulation method. It is straightforward to implement
it with, for example, the well known local Metropolis~\cite{METROPOLIS}
update algorithm. We can thus run simulations at constant~$\hat m$
and then combine them to obtain the canonical averages
with very high precision. It would be also possible 
to search for a cluster algorithm compatible with 
the tethered formalism, but we shall not investigate this issue here. 

One surprising feature of this Metropolis
implementation is that we can find no traces of critical
slowing down, within our errors, for all functions of $m$.  Other quantities,
such as the energy, exhibit the $z\approx 2$ behaviour
typical of local algorithms. Another interesting point
about this method is that, for a given temperature, 
$\varOmega_N(\hat m)$ has all the information about the system 
so we can, for example,
extract the canonical results at any value of the applied magnetic
field~$h$ without recomputing the 
tethered mean values (i.e., without any new simulations, see 
Section~\ref{results-h} below).

The TMC method is an extension of the 
strategy introduced in~\cite{MICROCANONICO}: there
the configuration space was extended to work in a microcanonical
ensemble, with entropy as the main physical variable.
The motivation in~\cite{MICROCANONICO} was handling first order
transitions without the need for tunnelling between two
coexisting phases. The microcanonical method was first applied in~\cite{MICROCANONICO}
to a pure system and was later, in~\cite{POTTS-PRIMER-ORDEN}, instrumental in simulating
a phase transition which remained first order in the presence of quenched 
disorder. Both the TMC and microcanonical methods
have deep links to Creutz's microcanonical demon~\cite{CREUTZ}. 
The main differences are: (i) we have as many demons as spins, 
(ii) our demons are continuous variables and (iii) we explicitly 
integrate out the demons, finding a tractable effective Hamiltonian. 

The effective potential can also be computed using multicanonical methods~\cite{MULTICANONICAL},
sometimes named multimagnetical~\cite{JANKE-CONDENSACION},
 or with the Wang-Landau algorithm~\cite{WANG-LANDAU}.
A major difference is that TMC does not require a random walk 
in magnetisation space (as with multimagnetical methods) or
in the energy-magnetisation plane (as in Wang-Landau).
Indeed, we have worked with as many as $10^6$
spins, while  $10^3$ spins is a typical limit for Wang-Landau
computations~\cite{WANG-LANDAU-2}.
On the other hand, standard micromagnetical methods~\cite{KAWASAKI}
do not render the effective potential.

In this paper we give a detailed exposition of the TMC method and demonstrate
it in the standard benchmark of the two dimensional Ising model. Our motivation
for this choice is twofold. On the one hand, since 
Onsager's solution~\cite{ONSAGER}  many other exact
results have been obtained (for a review see~\cite{WU-McCOY}
and references therein), which will help us check that our
answers are correct. This model is also the ideal scenario for cluster methods,
so even for those observables whose exact value is unknown we can 
check our results to a high degree of accuracy against a cluster simulation.  
On the other hand, the Ising model is sufficiently well known and simple
that we may concentrate on examining the details 
of the method with a minimum of nonessential discussion.
We purpose to show that TMC is capable of rendering very precise results
and try to convince the reader that it will still be efficient for harder problems.

The rest of the paper is organised as follows: 
\begin{itemize}
\item In Section~\ref{ensemble} we 
describe our statistical ensemble and its relationship 
with the standard canonical ensemble and properly define the effective potential.
In Section~\ref{method} we explain in detail how to set up a simulation using the 
TMC method. We then examine our own simulations for the Ising model. In Section~\ref{CSD}
we show that our algorithm presents no measurable traces of
critical slowing down for the magnetic field. 

\item We have carried out simulations at the critical point and in both the 
ferromagnetic and paramagnetic phases, with and without a magnetic field. 
Our results at the critical temperature and zero magnetic field
are presented in~\ref{results} and compared
both with the exact results at finite $L$ given by~\cite{FERDINAND-FISHER}
and with high precision Swendsen-Wang simulations (Section~\ref{results-Tc}).
In Section~\ref{peaks} we shall compute very accurately the magnetisation
critical exponent in a fairly unusual way, made easy by TMC.
In Section~\ref{results-paramagnetic} we check the performance 
of the method in the scaling paramagnetic region. Our chosen
test has been the computation of the first renormalised coupling constants.

\item The magnetic field $h$ is introduced in Section~\ref{results-h},
where we obtain several observables as a function of~$h$
and check our results with FSS arguments and by
recomputing the nonlinear susceptibilities with a 
finite differences formula. Section~\ref{results-ferromagnetic}
demonstrates the enhanced effectivity of the method
in the ferromagnetic regime, where we work
with and without an external field.
\item Finally, in Section~\ref{conclusions} we present our conclusions 
and outlook. We give some technical details of our numerical implementation
in an Appendix.
\end{itemize}

\section{The Statistical Ensemble}\label{ensemble}
\subsection{The model and the canonical observables}
We shall work with the $D=2$ Ising model, 
defined by the following partition function,
\begin{equation}\label{eq:Z}
Z = \sum_{\{\sigma_{\boldsymbol x}\}} \ee^{\beta \sum_{\langle \boldsymbol x, \boldsymbol y\rangle}
                                          \sigma_{\boldsymbol x} \sigma_{\boldsymbol y}
                                         + h \sum_{\boldsymbol x} \sigma_{\boldsymbol x}},
\qquad \sigma_{\boldsymbol x} = \pm 1,
\end{equation}
where $h$ is the applied magnetic field, $\boldsymbol x = (x_1,x_2)$ 
and $\medio{\boldsymbol x,\boldsymbol y}$ stands for
lattice nearest neighbours. We shall 
always consider square lattices with $N=L^2$ spins and periodic
boundary conditions. The infinite volume model has a second order phase transition
at a critical (inverse) temperature $\beta_\text{c}$ given by
\begin{equation}
\beta_\text{c} = \frac{\log(1+\sqrt{2})}2 = 0.440\,686\,793\,509\,771\ldots
\end{equation}
The simplest observables are
the energy and magnetisation,\footnote{We shall use sans-serif italics 
for random variables (i.e., functions of the spins) and serif italics for real 
numbers (e.g. expectation values or arguments
of the probability density functions). This will help emphasise 
which quantities are kept fixed and which can change.
We shall also use lowercase letters for intensive quantities and
uppercase letters for extensive quantities.}
\begin{align}
\aU &= N \au = - \sum_{\langle \boldsymbol x,\boldsymbol y\rangle} \sigma_{\boldsymbol x} \sigma_{\boldsymbol y},\\
\aM &= N \am = \sum_{\boldsymbol x} \sigma_{\boldsymbol x}.
\end{align}
In the canonical ensemble we are concerned with thermal averages, which we shall
denote by $\langle \ \cdot  \ \rangle_\beta$:
\begin{align}
U_\beta &= Nu_\beta = \langle \aU\rangle_\beta,\\
M_\beta &= Nm_\beta = \langle \aM\rangle_\beta.
\end{align}
The specific heat and magnetic susceptibility are
\begin{align}
C    &=  N [\langle \au^2\rangle_\beta - \langle \au\rangle^2_\beta],\label{eq:calor-especifico}\\ 
\chi_2 &=N [\langle \am^2\rangle_\beta-\langle \am\rangle_\beta^2].
\end{align}
The latter quantity can be seen as the zero momentum component of the two point
correlation function:
\begin{equation}\label{eq:G}
G_2(\boldsymbol k) = \frac1N \sum_{\boldsymbol x} \langle \sigma_{\boldsymbol x} \sigma_{\boldsymbol 0}\rangle_\beta\
                                          \  \ee^{\ii \boldsymbol k \cdot \boldsymbol x}.
\end{equation}
If we consider the asymptotic behaviour of the propagator in position space 
we arrive at the concept of \emph{correlation length},
\begin{equation}
\xi_\mathrm{exp} = \lim_{|\boldsymbol x| \to \infty} \frac{- |\boldsymbol x|}{\log \tilde G_2(\boldsymbol x)}\ . 
\end{equation}
This quantity is not easily measurable in our finite lattices, so we would like 
to have a better statistically behaved definition that could also be interpreted
as a correlation length. In momentum space
and in the limit $\xi_{\exp} |\boldsymbol k| \to 0$,
the propagator is well described by the free field form~\cite{AMIT-VICTOR,ZINN-JUSTIN}
\begin{equation}\label{eq:G-libre}
G_2(\boldsymbol k) \simeq \frac{A}{\xi^{-2} + 4 \sum_{i=1,2} \sin^2 k_i/2 } .
\end{equation}
($A$ is a constant). If we combine this formula at zero momentum and at the smallest 
nonzero momentum $\boldsymbol k_1$ we obtain
\begin{equation}\label{eq:xi}
\xi_1 = \frac{1}{2 \sin(\pi/L)} \left[ \frac{G_2(0)}{G_2(\boldsymbol k_{1})} - 1\right]^{1/2},
\end{equation}
with $G_2(\boldsymbol k_1)$ averaged over $\boldsymbol k_{1} = \left(2\pi/L, 0\right),\left(0,2\pi/L\right)$.
This definition~\cite{COOPER-xi} has proven
to be extremely useful in Finite Size Scaling studies~\cite{SOKAL-xi,UCM-xi}.

Equation~\eqref{eq:xi} does not work in the broken symmetry phase 
or with an applied magnetic field, because then
$G_2(\boldsymbol k)$ becomes singular at $\boldsymbol k=0$. We can still use~\eqref{eq:G-libre} for
$\boldsymbol k \neq 0$ and consider a second definition of the correlation length, 
now combining the two smallest nonzero momenta,
\begin{equation}\label{eq:xi2}
\xi_2 = \frac{1}{2 \sin(\pi/L)} 
\left[ \frac{G_2(\boldsymbol k_1)-G_2(\boldsymbol k_2)}
{2G_2(\boldsymbol k_{2}) -G_2(\boldsymbol k_1)}\right]^{1/2},
\end{equation} 
with $G_2(\boldsymbol k_2)$ averaged over  $\boldsymbol k_{2} =  \left({2\pi}/L, \pm {2\pi}/L\right)$
and $G_2(\boldsymbol k_1)$ as in definition~\eqref{eq:xi}.
The two definitions, $\xi_1$ and $\xi_2$, coincide for $\beta < \beta_\mathrm{c}$,
but only in thermodynamic limit.

Other observables are the Binder ratio
\begin{equation}
B = \frac{ \langle \aM^4\rangle_\beta}{\langle \aM^2\rangle^2_\beta}
\end{equation}
and the $2n$ point correlation functions at zero momentum,
\begin{equation}\label{eq:chi-2n}
\chi_{2n} = \frac1N \frac{\partial^{2n} \log(Z)}{(\partial h)^{2n}}.
\end{equation} 
Notice that these are just the cumulants of the magnetisation. In the low
temperature phase we should also consider odd derivatives.

\subsection{The tethered ensemble}
Let us consider the Ising model without an external field ($h=0$).
We can define a probability density function (pdf) for $\aM$, 
which will be a sum of $N+1$ Dirac deltas,
\begin{equation}\label{eq:p1}
p_1(m) = \frac1Z \sum_{\{\sigma_{\boldsymbol x}\}} \exp[-\beta \aU]
          \delta\biggl( m -\sum_i \sigma_i/N\biggr),\qquad
Z = \sum_{\{\sigma_{\boldsymbol x}\}} \exp[-\beta \aU].
\end{equation}
In the thermodynamic limit $p_1(m)$ is non vanishing for all $m$ in  $[-1,1]$.
We would like to have a new quantity that would
be continuous even for finite lattices. As a first step, 
we extend our configuration space with $N$ Gaussian demons:%
\footnote{This is not the only possible choice. In fact, 
we have also experimented with Poissonian demons,
better suited to a possible future implementation
of this method in dedicated computers with specific hardware~\cite{JANUS}.
The results (both in simulation time and accuracy of 
the final values) were virtually identical.} 
\begin{align}
Z &= \int_{-\infty}^\infty \prod_{i=1}^N \dd\eta_i\ 
       \sum_{\{\sigma_{\boldsymbol x}\}} \exp\biggl[-\beta \aU - \sum_i \eta_i^2/2\biggr],\qquad \aR = N\ar= \sum_i \eta_i^2/2.  \\
p_2(r) &= \frac1Z \int_{-\infty}^\infty \prod_{i=1}^N \dd\eta_i\
              \sum_{\{\sigma_{\boldsymbol x}\}} \exp\biggl[-\beta \aU - \sum_i \eta_i^2/2\biggr]
                  \delta\biggl(r - \sum_i \eta_i^2/(2N)\biggr).\label{eq:p2}
\end{align} 
Notice that in the canonical formalism the demons are a thermal bath decoupled 
from the spins because the spin contribution in~\eqref{eq:p2} cancels out.
By virtue of the Central Limit Theorem, the pdf for $\ar$ approaches
a Gaussian of mean $1/2$ and variance $(2N)^{-1}$ in the large $N$ limit.

Now we introduce $\hataM = N \hatam = \aM + \aR$.
The new variable $\hatam$ is continuous and its pdf is simply the convolution
of those of $\am$ and $\ar$ (as these are statistically independent):
\begin{equation}
p(\hat m) = \int_{-1}^1 \dd m\int_{0}^\infty\dd r\ p_1(m)p_2(r) \delta(\hat m- m -r).
\end{equation}
Notice that $\hataM \geq \aM$ and that $p(\hat m)$ is essentially a smoothed
version of $p_1(\hat m - 1/2)$. Finally, we introduce what will be our basic
physical quantity, the \emph{effective potential} $\varOmega_N(\hat m, \beta)$,
\begin{equation}\label{eq:p}
p(\hat m) = \exp[N\varOmega_N(\hat m, \beta)].
\end{equation}
The effective potential has all the information about the system 
at inverse temperature~$\beta$, including what would happen 
if it were immersed in an external magnetic field. 

Our next step is constructing the statistical ensemble and its relationship
with the canonical one. In order to do this we represent $\varOmega_N(\hat m,\beta)$
as a functional integral and integrate out the demons,
\begin{align}
\ee^{N\varOmega_N(\hat m,\beta)}
  &= \frac1Z \int_{-\infty}^\infty \prod_{i=1}^N\dd\eta_i \sum_{ \{\sigma_{\boldsymbol x}\}}
      \ee^{-\beta\aU-\sum_i\eta_i^2/2} \delta\biggl(\hat m-\am-\sum_i \eta^2_i/(2N)\biggr)\nonumber\\
  &= \frac1Z\int_{-\infty}^\infty \prod_{i=1}^N\dd\eta_i \sum_{ \{\sigma_{\boldsymbol x}\}}
      \ee^{-\beta \aU + \aM-N\hat m }\delta\biggl(\hat m- \am-\sum_i \eta^2_i/(2N)\biggr)\nonumber\\
  &=\frac1Z \sum_{\{\sigma_{\boldsymbol x}\}} \ee^{-\beta \aU +\aM- N\hat m }
     (\hat m -\am)^{(N-2)/2} \frac{(2\pi N)^{N/2} \theta(\hat m -\am)}{\varGamma(N/2)} \label{eq:exp-Omega}
\end{align}
The condition $\hatam \geq \am$ is explicitly enforced by the Heaviside step function~$\theta$.
Notice as well that we have constructed the effective potential
from the starting point of a canonical ensemble. 
It would be elementary to retrace our steps for a microcanonical $\varOmega_N$ (we 
would just have to change the exponential of the energy in the previous equations to the
appropriate microcanonical weight, see~\cite{MICROCANONICO}).

We want to develop a statistical ensemble based on the effective potential 
(i.e. to define average values). To do this we differentiate $\varOmega_N$, 
\begin{equation}\label{eq:der-Omega}
\frac{\partial \varOmega_N(\hat m,\beta)} {\partial \hat m} = 
\frac{\sum_{\{\sigma_{\boldsymbol x}\}}  \left(-1 + \frac{N-2}{2N(\hat m-\am)}\right)
                                          \omega(\beta,\hat m, N; \{\sigma_{\boldsymbol x}\})}
                                    {\sum_{\{\sigma_{\boldsymbol x}\}} \omega(\beta, \hat m, N; \{\sigma_{\boldsymbol x}\})} .
\end{equation}
where
\begin{equation}\label{eq:omega}
\omega(\beta,\hat m, N;\{\sigma_{\boldsymbol x}\}) = \ee^{-\beta \aU +\aM- N\hat m} (\hat m -\am)^{(N-2)/2}
                                    \ \theta(\hat m -\am).
\end{equation}
Equations~\eqref{eq:der-Omega} and~\eqref{eq:omega} suggest a new ensemble
where the probability of a given configuration $\{\sigma_{\boldsymbol x}\}$
is proportional to~$\omega( \beta,\hat m,N; \{\sigma_{\boldsymbol n}\})$. 
Therefore, we can define the \emph{tethered mean value} $\medio{\ \cdot\ }_{\hat m,\beta}$
for an arbitrary observable $\aO$ by
\begin{equation}\label{eq:medio-O}
\langle \aO \rangle_{\hat m,\beta} = 
  \frac{ \sum_{\{\sigma_{\boldsymbol x}\}} \aO(\hat m; \{\sigma_{\boldsymbol x}\}) \omega(\beta,\hat m, N; \{\sigma_{\boldsymbol x}\})}
       {\sum_{\{\sigma_{\boldsymbol x}\}} \omega(\beta, \hat m, N; \{\sigma_{\boldsymbol x}\})}\, .
\end{equation}
Now we define the tethered magnetic field as
\begin{equation}
\hatah(\hat m;\{\sigma_{\boldsymbol x}\} ) = -1 + \frac{N/2-1}{\hat M-\aM}
\end{equation}
and notice from \eqref{eq:der-Omega} and~\eqref{eq:medio-O} that 
\begin{equation}
\langle \hatah \rangle_{\hat m,\beta} = \frac{\partial \varOmega_N(\hat m,\beta)}
                                        {\partial \hat m}.
\end{equation}

The duality between the roles of $\hatah$ and $\am$ in the canonical
and tethered ensembles is best illustrated from 
the tethered fluctuation-dissipation formula:
\begin{equation}
\frac{\partial \medio{\aO}_{\hat m,\beta}}{\partial \hat m} =
\left\langle\frac{\partial \aO}{\partial \hat m}\right\rangle_{\hat m,\beta}+
N \left[ \medio{\aO\hatah}_{\hat m,\beta} - \medio{\aO}_{\hat m,\beta} \medio{\hatah}_{\hat m,\beta}
\right].
\end{equation}

The simulation strategy is then clear: we compute the tethered averages 
of $\hatah$ and whichever observables we need for a reasonable number
of values of $\hat m$ (we shall discuss what we mean by `reasonable' in the
next section). The effective potential cannot be measured directly, 
but we do have its derivative~$\medio{\hatah}_{\hat m,\beta}$. Integrating~$\langle \hatah\rangle_{\hat m,\beta}$ and 
requiring that the probability $p(\hat m)$ be normalised we 
can obtain $\varOmega_N(\hat m, \beta)$ unambiguously.

Once we have the effective potential, we can recover the  canonical
averages with the following formula:
\begin{equation}\label{eq:canonical-average}
\langle \aO\rangle_\beta = \int\dd \hat m\ \langle \aO\rangle_{\hat m,\beta}
                              \ \ee^{N \varOmega(\hat m,\beta)}\ .
\end{equation}
Thus far, we have defined the ensemble in the absence of an external
magnetic field $h$, but we can include it very easily.
Indeed, we just have to notice that an applied field introduces a 
shift in the origin of $\hatah$. The computation of canonical averages
with an external field is then straightforward, using the same tethered mean
values we had for $h=0$:
\begin{equation}\label{eq:canonical-average-h}
\langle \aO \rangle_{\beta}(h) = \frac{\int\dd \hat m\ \ee^{N[\varOmega_N(\hat m,\beta) + h\hat m]}
\medio{\aO}_{\hat m,\beta}}{\int\dd \hat m\ \ee^{N[\varOmega_N(\hat m,\beta) + h\hat m]} } .
\end{equation}
The denominator is necessary because now the shifted effective potential
is not normalised. 

\section{Description of the simulations}\label{method}
The basic steps in a TMC simulation, which we shall discuss later,
are the following.
\begin{enumerate}
\item Select an appropriate sampling for $\hataM$ remembering
that, essentially, $\hataM \simeq  \aM + N/2$. We shall discuss
the choice of the $\hat m$ grid in Section~\ref{sampling}. Naturally, we must 
restrict ourselves to a finite simulation range, $[\hat m_\text{min},\hat m_\text{max}]$,
which introduces an (exponentially small) cutoff error. 

\item Run independent simulations for each $\hat m$, measuring
the tethered averages of $\hatah$ and the other relevant observables.
\item The mean values $\langle \aO\rangle_{\hat m,\beta}$ are
smooth functions of $\hat m$, so they can be interpolated
safely. We use cubic splines~\cite{NR}, but other methods may also work.
Appendix~\ref{detalles-tecnicos} has some technical details
about this and other points of our implementation.
\item Integrate $\langle \hatah\rangle_{\hat m,\beta}$ for the whole range
of $\hat m$. We use an average of the integral in both 
directions to reduce the systematic error:
\begin{equation}\label{eq:IN}
I_N(\hat m,\beta) = \frac12 \left(\int_{\hat m_\text{min}}^{\hat m} \dd \hat m'\ 
              \langle \hatah\rangle_{\hat m',\beta} 
             - \int_{\hat m}^{\hat m_\text{max}} \dd \hat m'\ 
              \langle \hatah\rangle_{\hat m',\beta}\right).
\end{equation}
This defines $\Omega_N(\hat m, \beta)$ up to an additive constant.
Notice that this is not the same as forcing $\varOmega_N(\hat m,\beta)$
to be symmetric (which it cannot be, since $\hatam$ has a finite lower
bound but extends to infinity).
\item Normalise the pdf,
\begin{equation}
c = \int_{\hat m_\text{min}}^{\hat m_\text{max}}\dd\hat m\ \exp[N I_N(\hat m,\beta)].
\end{equation}
Then the effective potential is
\begin{equation}
\varOmega_N(\hat m, \beta) = I_N(\hat m, \beta) - \frac1N\log c.
\end{equation}
\item Compute the canonical averages with the interpolated 
tethered averages and equation~\eqref{eq:canonical-average}. The statistical
errors are easily estimated with the jackknife method~(see, e.g., \cite{AMIT-VICTOR}).
The $i$th block of our splines interpolates the $i$th jackknife blocks of each simulated
tethered mean value.
\item To obtain canonical results with a magnetic field $h$, simply substitute $\varOmega_N$
with $\varOmega_N^h$,
\begin{equation}\label{eq:Omega-h}
\varOmega_N^h(\hat m,\beta, h) = \varOmega_N(\hat m, \beta) + h\hat m -\frac1N\log c',
\end{equation} 
where $c'$ is the new normalisation constant and use 
\begin{equation}\label{eq:valor-canonico-h}
\medio{\aO}_{\beta}(h) = \int \dd\hat m \ \ee^{N \varOmega_N^h(\hat m,\beta,h)} \medio{\aO}_{\hat m, \beta}.
\end{equation}
\end{enumerate}
\begin{figure}[t]
\centering
\includegraphics[height=.7\linewidth,angle=270]{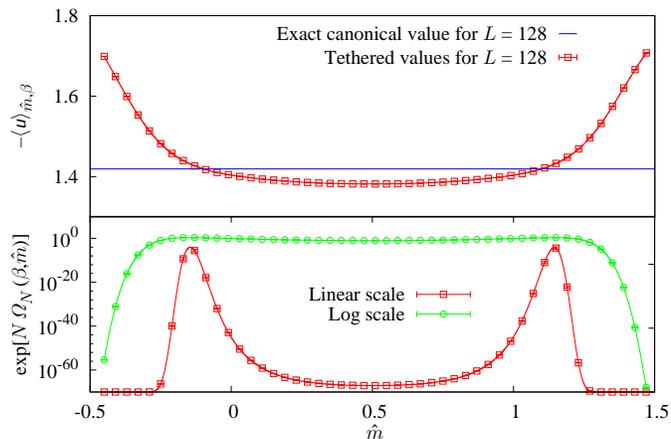}
\caption{Computation of the canonical average from tethered averages at
the critical temperature. The lower panel depicts the pdf of $\hatam$,
in logarithmic (left axis) and linear (right axis) scales,
so that both the peaks and the tails can be seen. 
On the upper panel we plot the tethered mean values
$-\medio{\au}_{\hat m,\beta_\mathrm{c}}$ as a function of $\hat m$. 
The horizontal line is the exact value for $L=128$ computed from~\cite{FERDINAND-FISHER},
$\medio{\au}_{\beta_\mathrm{c}} = - 1.419\,076\ldots$
The continuous curves are our cubic spline interpolations (see Appendix).
Our final result is $\langle \au\rangle_{\beta_\mathrm{c}}= -1.419\,05(5)$.
Notice that the range of tethered values for $\medio{\au}_{\hat m,\beta_\mathrm{c}}$
is several orders of magnitude greater than our statistical error.
}
\label{fig:tethered}
\end{figure}
Fig.~\ref{fig:tethered} pictures this process. On the top panel we 
represent the energy density as a function of $\hat m$, together 
with a horizontal line marking the canonical average at the
critical point. On the bottom panel we plot the pdf 
$p(\hat m)$. This function is reconstructed with great precision (we have to
keep in mind that there is a factor $N=L^2$ in the exponent).
The rendered accuracy in $p(\hat m)$ allows us to obtain 
$\langle \au\rangle_{\beta_\mathrm{c}} = -1.419\,05(5)$, correct
to five significant figures, even though the range of 
variation of $\langle \au\rangle_{\hat m}$ is of 
$\sim 20\%$. The graphs were generated from the simulations
described in Section~\ref{parameters}.

Once the values of $\hat m$ have been selected, the independent
simulations are carried out in a standard way. We use 
Metropolis dynamics to update the configuration.
Let $\{\sigma_{\boldsymbol x}\}$ be the 
initial configuration and $\{\sigma_{\boldsymbol x}'\}$
the proposed new configuration (where one of the spins 
has been reversed). Then the probability of 
accepting the change is, from~\eqref{eq:omega},
\begin{equation}\label{eq:Metropolis}
P\bigl(\{\sigma_{\boldsymbol x}\}\to\{\sigma_{\boldsymbol x}'\}\bigr) =
\begin{cases}0, & \text{if $\aM' > \hat M$}\\
\min\left\{1, \ee^{\Delta\textsf{\itshape S}}\right\}, & \text{if $\aM' \leq \hat M$},
\end{cases}
\end{equation}
where
\begin{equation}\label{eq:deltaAccion}
\Delta \textsf{\itshape S} = -\beta(\aU'-\aU) + (\aM'-\aM)+ \left(\frac N2 -1\right)
                              \log\left(1-\frac{\aM'-\aM}{\hat M-\aM}\right)\, .
\end{equation}

An interesting point about our algorithm is that 
it is `compulsory parallel'. The fact that we have to run
simulations at several values of $\hat{m}$ adds
a layer of trivial parallelisation: we simply perform the simulations
for each~$\hat m$ independently and only later combine
their results to construct $\varOmega_N$ (an operation that 
is essentially costless in computer time).

A cautionary note: as our pseudorandom number generator we were originally using a 
64 bit congruential generator reported in~\cite{OSSOLA-SOKAL}.
We simulated the lattice sequentially and extracted
a random number per site, although in principle we would only
need one when $\Delta \textsf{\itshape S}<0$ in~\eqref{eq:deltaAccion}. 
This matches the conditions studied in~\cite{OSSOLA-SOKAL,OSSOLA-SOKAL-2}, where
significant deviations from the expected values were found
using the same generator for a $D=3$ Ising model. The
error showed up only for large lattices, $N\geq 128^3$. 
Perhaps unsurprisingly, we also obtained wrong results (farther than $3$ 
standard deviations from the exact values) 
for a system with a similar number of spins ($N=1024^2$).
This is our biggest system and we did not have any problems for 
all the smaller ones. Once this issue was identified, we 
added to the congruential generator a 64 bit version 
of the shift register method reported in~\cite{PARISI-RAPUANO}
and redid all our computations.
All the numerical results presented 
in this paper have been obtained using
the sum of both generators (modulo $2^{64}$).

\subsection{How to select a good sampling of $\hat m$}\label{sampling}
A good choice of the $\hat m$ we are going to simulate may reduce
systematic and statistical errors significantly. Remember that while $-1\leq m \leq 1$,
$\hat m$ in principle may extend to infinity. Actually, 
$p(\hat m)$ has completely negligible values outside the range 
$[-1/2,3/2]$, so we can restrict ourselves to that interval. 
If we look at Fig.~\ref{fig:tethered}
we see that $p(\hat m)$ is a two peaked distribution,
so values of $\hat m$ inside the peaks contribute more that
those in the middle or in the tails. These peaks get 
closer together and slightly narrower as we increase $L$, so
it may seem that the choice of $\hat m$ is quite delicate (specially
considering we do not know $p(\hat m)$ until we have run our simulation).%
\footnote{The discussion in this Section is relevant to the disordered 
phase and the critical region. In the broken symmetry phase, 
the peaks rapidly get very high
and extremely narrow as we increase $L$ and the criteria
 are different (see Section~\ref{results-ferromagnetic}).} 
A different, but related, question also arises: is it better to
compute $\medio{\hatah}_{\hat m,\beta}$ at 
more points or more precisely with a coarser grid? We shall try to give
some practical recipes below. 

The question is easy to analise for $I_N$, see equation \eqref{eq:IN}. Let us
assume that we have obtained the mean value of $\hatah$ at $K$
points in a grid. Our estimator $[\hatah]_k$ 
is related to the actual value by
\begin{equation}
[\hatah]_k = \medio{\hatah}_k + \eta_k, \qquad k = 1,\ldots,K,
\end{equation}
where $\eta_k$ are the errors, expected to be Gaussian distributed, of
similar size and statistically uncorrelated. Hence, our numerical 
estimate for $I_N$ will be given by a quadrature formula
\begin{equation}
I_N \simeq \sum_{k=1}^K g_k \medio{\hatah}_k + \sum_{k=1}^K g_k \eta_k.
\end{equation}
In this equation the first summand is subject to 
systematic error while the second one is the statistical error.
Now, since the quadrature coefficients $g_k$ scale as $1/K$,
it is clear that the statistical
error will scale as $1/\sqrt{K}$. This suggests that doubling
the number of points of the grid is equivalent to doubling
the simulation time for each one. However, the analysis
for canonical mean values is fairly more involved, since
the errors in $\varOmega_N$ will be highly correlated
for different $\hat m$. Therefore, we perform a numerical 
experiment.

\begin{table}
\centering
\caption{Final value for $-\langle \au\rangle_{\beta_\mathrm{c}}$ as we
change the number of points for the reconstruction of $\varOmega_N$ 
and their precision. MCS = Monte Carlo Sweeps for each point. 
The runs labelled `uniform sampling' consist of $N_\text{points}$
values of $\hat m$ evenly distributed in the range $[-0.4,1.4]$.
The runs labelled `improved sampling' have $2/3N_\text{points}$ points evenly
distributed in the same range, plus and additional $N_\text{points}/3$ 
inside the peaks, effectively doubling the density in the dominant regions.
The last results of each column have a similar precision, but those
computed with uniform sampling required twice the simulation time.}
\smallskip
\label{tab:puntos}
\begin{tabular*}{\columnwidth}{@{\extracolsep{\fill}}lllll}
\cline{2-5}
& \multicolumn{2}{c}{\bfseries Uniform sampling} 
& \multicolumn{2}{c}{\bfseries Improved sampling}\\
\hline
 \multicolumn{1}{c}{\boldmath $N_\mathrm{points}$} & \multicolumn{1}{c}{\bfseries \boldmath $10^6$ MCS} &\multicolumn{1}{c}{\bfseries \boldmath $10^7$ MCS}&
 \multicolumn{1}{c}{\bfseries \boldmath $10^6$ MCS} &\multicolumn{1}{c}{ \bfseries \boldmath $10^7$ MCS}\\
\hline
12    & 1.437\,28(33)  &1.437\,38(11)   &                &\\               
23    & 1.419\,25(22)  &1.419\,117(6)   &                & \\               
46    & 1.419\,21(13)  &1.419\,117(43)  & 1.419\,08(11)  & 1.419\,107(38)   \\
91    & 1.419\,14(10)  &1.419\,093(36)  & 1.419\,13(8)   & {\bf 1.419\,128(31)}\\
181   & 1.419\,14(7)   &{\bf 1.419\,095(28)}  & 1.419\,034(5)  & \\ 
451   & 1.419\,06(5)   &                & 1.419\,073(39) & \\       
901   & 1.419\,065(33) &                & {\bf 1.419\,077(26)} &   \\
1801  & {\bf 1.419\,062(24)}                   \\   
Exact & \multicolumn{4}{c}{1.419\,076\,272\,086\dots}\\
\hline
\end{tabular*}
\end{table}

Table~\ref{tab:puntos} shows the values for the energy 
density at the critical point, $-\langle \au\rangle_{\beta_\mathrm{c}}$,
obtained in different series of runs. In the first column, 
we use evenly spaced points with $10^6$ Monte Carlo
Sweeps (MCS) each. In the second column the points
are also uniformly distributed, but now we perform
$10^7$ MCS in each of them. The third and fourth
columns are analogous, but with a greater density of points
inside the peaks. We can reach several conclusions from this
table:
\begin{itemize}
\item If we use too low a number of points we will see a significant
systematic error, no matter how precise they are. 
\item Once the systematic error is under control, increasing the 
number of evenly distributed points has an effect of $1/\sqrt{N_\mathrm{points}}$
in the statistical error. This is best seen in the leftmost column.
The effect is roughly the same if we increase
the number of MCS in each point by the same factor (the errors
of the first column are $\sim\sqrt{10}$ times greater than the
corresponding ones of the second).
\item If we add more points inside the peaks, the error
may decrease faster than $1/\sqrt{N_\mathrm{points}}$. 
The errors with $N_\text{points}$ and uniform sampling are 
only slightly smaller than those with $N_\text{points}/2$ 
and improved sampling.
\end{itemize}
We can summarise this analysis with the following prescription
for the choice of $\hat m$:
\begin{enumerate}
\item Run a first simulation with enough uniformly sampled $\hat m$ 
so that the systematic error is small or unnoticeable (i.e., so
that the peaks of the distribution can be roughly reconstructed
and the tails are reliably sampled).
We have used $\sim 50$. This may seem a big number, but we must
take into account that we have only looked at the energy 
in Table~\ref{tab:puntos}. Other quantities, such as high 
moments of the magnetisation or observables at a nonzero
magnetic field, require that the tails of the distribution
be reasonably well sampled.
\item Add a similar number of points inside the peaks of the pdf to 
eliminate the systematic error and reduce the statistical error.
\end{enumerate}
We have found that the second step is not always necessary. In fact, 
for lattices up to $L=256$ we have limited ourselves to 
computing $51$ evenly distributed $\hat m$. For bigger
lattices the peaks are steeper and we have added
another $26$ points inside them. 

\subsection{Other practical recipes}\label{momentos}
It is sometimes interesting to compute
high moments of the magnetisation (for example,
see Section~\ref{results-paramagnetic}). One obvious possibility
is to simply measure individual values
for $\am^\ell(\hat m; \{\sigma_{\boldsymbol x}\})$ and then compute the 
tethered and canonical averages as usual. But our
method provides an alternative way of calculating
$\langle \am^\ell\rangle_\beta$. Indeed, we have 
the whole pdf $p(\hat m)$ and we know
that $\hataM = \aM + \aR$. Now, the moments
for $\aR$  can be easily obtained analytically, so
it suffices to compute $\langle \hatam^\ell\rangle_\beta$ 
to reconstruct $\langle \am^\ell\rangle_\beta$ without 
any need for individual measurements of $\am^\ell$.
For example,
\begin{align}
\langle \am^2\rangle_\beta &= \langle (\hatam-1/2)^2\rangle_\beta - \frac{1}{2N}\ .\\
\langle \am^4\rangle_\beta &= \langle (\hatam-1/2)^4\rangle_\beta
                              -\frac{3}{N} \langle \am^2\rangle_\beta 
                              - \frac{3}{4N^2}+\frac{3}{N^3}\ .
\end{align}
These formulas are valid for the symmetric phase, where $\langle \am\rangle_\beta = 0$. We have computed the moments of the magnetisation up to
$\langle \am^8\rangle_\beta$ both from individual measurements and with this
procedure and the results are identical. This will not 
be at all surprising once we see Section~\ref{CSD}, where
it is shown that the correlation time for $\langle\am\rangle_{\hat m,\beta}$
is $<1$ (which means that the uncertainty in $p(\hat m)$ is 
going to determine the total error). 

\section{Critical slowing down}\label{CSD}
\begin{figure}[b]
\centering
\includegraphics[height=.75\linewidth,angle=270]{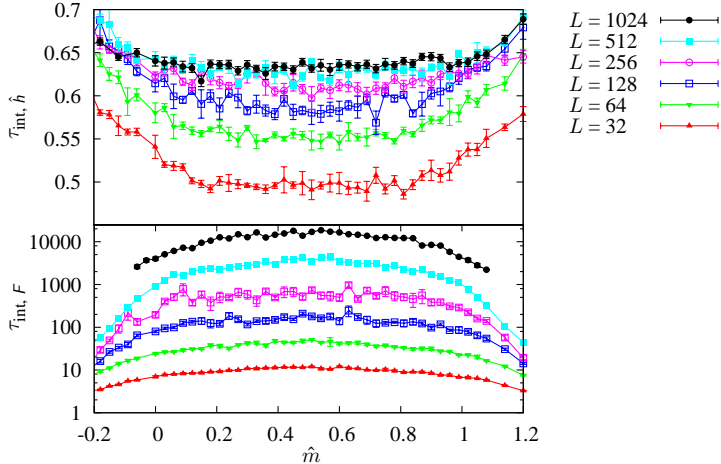}
\caption{Integrated autocorrelation time for $\hat h$ and $F$
as a function of $\hat m$, computed with
the self consistent window method~\cite{SOKAL-TAU}. Notice the absence
of critical slowing down for the former. In the lower panel,
as the correlation time for $F$ grows as $L^2$, we did not
use individual measurements of $F$ after each Monte Carlo step
for $L\geq512$. Instead, we averaged over $50$ such measurements for $L=512$ 
and over $4000$ measurements for $L=1024$.
We then computed the autocorrelation functions
for these blocked measurements (of course, we multiplied the 
integrated times thus obtained by the length of the blocks). This 
accounts for the smaller errors in $\tau_{\text{int}}$ for these lattices.
For $L=1024$ the correlation time becomes unmeasurable (i.e., smaller than our blocks)
for $\hat m >1.1$ or $\hat m<-0.1$.
} 
\label{fig:tau_int}
\end{figure}
Before we examine our results, we must make sure that we
have been able to thermalise our systems and that we have enough
independent measurements to generate precise averages. We would also like
to know whether our algorithm suffers from critical slowing down (CSD) and, if so,
in what measure.

To address these questions we have computed the autocorrelation
functions and integrated autocorrelation times~\cite{SOKAL-TAU,AMIT-VICTOR} for $\au$, $\hatah$, $\am$ and for each value 
of $\hat m$ (we will restrict ourselves to the critical temperature
in this section). Another observable will be of interest. Remembering definitions~\eqref{eq:G}
and~\eqref{eq:xi}, we can introduce the tethered mean value of the two point propagator
at the smallest momentum:
\begin{equation}\label{eq:F}
\medio{\aF}_{\hat m,\beta} = \frac1N \sum_{\boldsymbol n} \medio{\sigma_{\boldsymbol x} \sigma_{\boldsymbol 0}}_{\hat m,\beta} \ee^{\ii \boldsymbol k_1 \cdot \boldsymbol x}.
\end{equation}

We define the autocorrelation function at time $t$ for an observable $\aO$ by
\begin{equation}
C_\aO(t,\hat m) = \langle \aO_s \aO_{s+t}\rangle_{\hat m} - \langle \aO\rangle_{\hat m}^2,\qquad \rho_\aO(t,\hat m) = \frac{C_\aO(t,\hat m)}{C_\aO(0,\hat m)}\ .
\end{equation}
From this function we can obtain several characteristic times~\cite{SOKAL-TAU}:
\begin{align}
\tau_{\mathrm{exp}, \aO}(\hat m) &= \lim_{t\to\infty}\sup \frac{t}{-\log|\rho_\aO(t,\hat m)|},& &\text{exponential time of $\aO$},\\
\tau_{\mathrm{exp}} &= \sup_{\aO}\ \tau_{\mathrm{exp},\aO},& &\text{exponential time of the system},\\
\tau_{\mathrm{int}, \aO}(\hat m) &= \frac12 + \sum_{t=1}^\infty \rho_\aO(t,\hat m),& &\text{integrated time of $\aO$}.
\end{align}
The first two measure the amount of time (i.e. number of Monte Carlo sweeps) that must pass before
the system is thermalised. The third characterises the minimum time difference so that 
two measurements can be considered independent (i.e. uncorrelated). 
\begin{figure}[t]  
\centering
\includegraphics[height=.7\linewidth,angle=270]{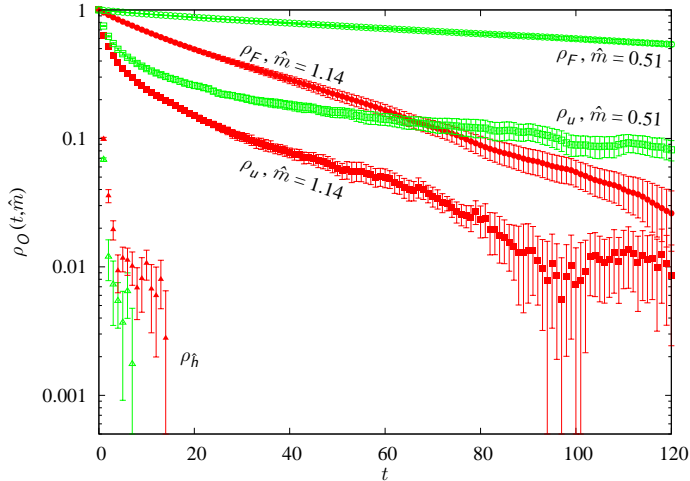}
\caption{Normalised autocorrelation functions of $\au$, $\aF$ and $\hatah$ 
at the minimum ($\hat m=0.51$, black symbols) and one of the maxima
($\hat m = 1.14$, grey symbols) of $p(\hat m)$ for $L=128$
at the critical temperature. We appreciate how the curves for $\au$ (circles) and $\aF$ (squares)
become parallel after a small number of steps. We have only plotted $\rho_{\hatah}$ (triangles) until
the point where it first becomes compatible with zero, to avoid cluttering the graph.}
\label{fig:corr}
\end{figure}

We compute $\tau_{\mathrm{int},\aO}$ using the self consistent window
method~\cite{SOKAL-TAU}.
In Fig.~\ref{fig:tau_int} we show the integrated autocorrelation times for $\aF$ and $\hatah$ as a function
of~$\hat m$. 

We see a clear difference: while the time scales for $\aF$ grow as $L^z$, with $z\approx2$;
the $\tau_{\mathrm{int}}$ for $\hatah$ do not exhibit any significant increase. That is, even though
we are using a local algorithm, some of our observables do not experience any critical slowing down
(and $\hat h$ is a particularly important observable, as we integrate it to obtain
the effective potential). The integrated times for the energy are much smaller
than those of $\aF$, but their behaviour is qualitatively similar (i.e., they grow
as $L^2$). Within the computed quantities, $\aF$ is the slowest mode
for our algorithm. In some sense, the absence of CSD for all functions of $\am$
(including $\hatah$) 
is not completely surprising, because the Metropolis update uses nonlocal
information that involves the whole system through $\am$.

The exponential times are much more difficult to compute precisely, but we can easily
give a rough estimate. Fig.~\ref{fig:corr} shows the normalised autocorrelation functions
for the energy and $\aF$ at two different values of $\hat m$ (one at one of 
the maxima of $p(\hat m)$, the other at the minimum). We see that $\aF$ is a rather
pure mode for our Metropolis dynamics, with a very clean exponential decay for small times, so that 
$\tau_{\mathrm{exp},\aF} \approx \tau_{\mathrm{int}, \aF}$. The curve for the energy
falls rapidly at first and then becomes parallel to $\rho_\aF(t,\hat m)$.
This is a clear indication that the integrated time for $\aF$ can be considered
as a good approximation to the exponential (i.e. thermalisation) time of the 
system.

The reader may find this section in conflict with the exact result by 
Hohenberg and Halperin~\cite{HOHENBERG-HALPERIN}, who showed
that for conserved order parameter dynamics (model $B$) in 
Ising models the critical exponent is $z=4-\eta$. Nevertheless, even our 
slowest mode has a much smaller $z$. The way out of this paradox 
is in the nonlocal nature of our conservation law. Model $B$
needs magnetisation diffusion across the boundary if the 
magnetisation is to change in the enclosed region. When the locality
constraint is violated, the new dynamical exponent is expected to be 
$z_\text{nonlocal}=z-2$~\cite{TAMAYO-KLEIN}. Clearly, this is the case
for our dynamics, where a change of $\hat m$ inside a given region
does not occur through diffusion across its boundary.

\section{Results at \boldmath$\beta_\mathrm{c}$, $h=0$}\label{results}
We summarise here our results at the critical point, which
correspond to the bulk of our simulations. 
This section is divided in three parts. First we 
provide the parameters of our simulations. Second,
we report our results for the mean values that would be considered
in a traditional canonical computation. 
Finally, in subsection~\ref{peaks} we perform a rather unconventional,
but very accurate, computation of the critical exponents ratio $\beta/\nu$
taking advantage of the peculiarities of the tethered formalism.

\subsection{Parameters of our simulations}\label{parameters}
We have simulated 
lattice sizes $L=16,32,\ldots,1024$.  
For systems with $L\leq 256$ we have used $51$ uniformly distributed
values of $\hat m$ in $[-0.5,1.5]$ (except for $L=16$, where
we had to extend the range to $[-1.61,1.61]$ to avoid cutoff errors).
As we shall see in Table~\ref{tab:results-Tc},
this choice (considered suboptimal in Section~\ref{method}) does not
introduce any discernible systematic error.  We have been very conservative
and chosen a wider range for $\hat m$ than what we used in Section~\ref{sampling},
slightly enlarging the computations to avoid any chance of a cutoff error.
For $L\geq512$ we ran a first simulation with $51$ equally spaced~$\hat m$
in $[-0.3,1.3]$. In these cases the peaks were considerable narrower and closer
together than for the smaller lattices. To avoid any discretisation errors,
we have followed the practical
recipe given in Section~\ref{method}: after obtaining an 
approximation to the shape of $\varOmega_N$, we have added 
another $26$ points, doubling the density inside the peaks. 

In all cases we have performed $10^7$ Metropolis sweeps
for each value of~$\hat m$. Following the standard prescription~\cite{SOKAL-TAU},
we have discarded a fifth of the 
measurements for thermalisation (although the correlation 
times are much smaller, see~Section~\ref{CSD}) and 
formed $100$ jackknife blocks~\cite{AMIT-VICTOR} with 
the rest to obtain a reliable estimate of the statistical
error. According to Section~\ref{CSD} and, particularly, 
Fig.~\ref{fig:tau_int}, this means that for $L=1024$
the length of the blocks is only about  $5$ times
the largest found $\tau_\text{exp}$ (that of $\hat m=0.5$).
There has been no need to increment the numerical effort in
this point as this exponential time corresponds to 
the minimum of $p(\hat m)$ and we have seen that the canonical values
are dominated by the neighbourhood of the peaks.
There the exponential time is an order of magnitude smaller,
so our error estimates are sound (as can be seen
by comparing our results with the exact values).
 In any case, we have recomputed
the errors of Table~\ref{tab:results-Tc}
with $50$ and $200$ jackknife blocks and checked that they remain completely stable.
Our canonical simulations consisted of  $10^7$  Swendsen-Wang updates.

\begin{table}[b]
\centering
\caption{Results at the critical temperature and comparison with a cluster algorithm. (T): Tethered Monte Carlo,
(C): Swendsen-Wang, (E): Exact results at finite $L$ from~\cite{FERDINAND-FISHER}.}
\label{tab:results-Tc}
\smallskip

\begin{tabular*}{\columnwidth}{@{\extracolsep{\fill}}llllllll}
\hline
\boldmath $L$ & \multicolumn{1}{c}{\boldmath $-\medio{\au}_{\beta_\mathrm{c}}$} &
\multicolumn{1}{c}{\boldmath $\chi_2/L^2$} & \multicolumn{1}{c}{ \boldmath $\xi_1/L$} & 
\multicolumn{1}{c}{\boldmath $\xi_2/L$} &\multicolumn{1}{c}{\boldmath $C$}  &
\multicolumn{1}{c}{\boldmath $B$} & \multicolumn{1}{c}{\boldmath $\partial_\beta\xi_1/10^4$} \\
\hline
16 (T)        & 1.453\,08(4) & 0.545\,43(6) & 0.911\,6(2)  & 0.246\,13(13)& 7.718\,6(14) & 1.165\,62(7) & 0.036\,547(9)\\
16 (C)        & 1.452\,9(2)  & 0.545\,1(3)  & 0.910\,4(9)  &              & 7.718(10)    & 1.165\,9(3)  & 0.036\,50(6)\\
16 (E)        & 1.453\,065\ldots&           &              &              & 7.717\,134\ldots               \\
\hline
32 (T)        & 1.433\,69(4) & 0.459\,00(10)& 0.907\,2(4)  & 0.242\,2(3)  & 9.509(3)   & 1.167\,23(14) & 0.144\,07(7)\\
32 (C)        & 1.433\,67(12)& 0.459\,1(2)  & 0.907\,8(9)  &              & 9.493(13)  & 1.167\,1(3)   & 0.144\,1(3) \\
32 (E)        & 1.433\,659\ldots&           &              &              & 9.509\,379\ldots\\
\hline
64 (T)       & 1.423\,97(4) & 0.386\,19(18) & 0.906\,5(9)   & 0.240\,0(5)  & 11.285(6)  & 1.167\,5(3)   & 0.573\,8(6)\\
64 (C)       & 1.423\,90(6) & 0.386\,0(2)   & 0.905\,6(10)  &              & 11.293(17) & 1.167\,7(4)   & 0.573\,1(11)\\
64 (E)       & 1.423\,938\ldots &           &               &              & 11.288\,138\ldots\\          
\hline 
128 (T)      & 1.419\,05(5) & 0.324\,4(3)   & 0.904\,0(18)  & 0.240\,8(11) & 13.063(10) & 1.168\,4(7)   & 2.289(5)\\
128 (C)      & 1.419\,06(4) & 0.324\,59(17) & 0.904\,8(10)  &              & 13.06(2)   & 1.167\,7(4)   & 2.287(4)\\
128 (E)      & 1.419\,076\ldots &           &               &              & 13.060\,079\ldots \\ 
\hline
256 (T)      & 1.416\,63(5) & 0.272\,8(6)   & 0.904(4)      & 0.240(2) & 14.83(2) & 1.168\,7(14) & 9.16(4)\\
256 (C)      & 1.416\,64(2) & 0.272\,86(14) & 0.904\,2(9)     &          & 14.83(2) & 1.168\,2(4)  & 9.14(2) \\
256 (E)      & 1.416\,645\ldots&            &               &          & 14.828\,595\ldots \\
\hline
512 (T)      & 1.415\,42(4)  & 0.229\,3(7)  & 0.903(6)      & 0.240(4) & 16.57(3) & 1.168(2)     & 36.38(19)\\ 
512 (C)      & 1.415\,444(11)& 0.229\,68(13)& 0.905\,9(10)  &          & 16.60(2) & 1.167\,6(4)    & 36.64(10)\\
512 (E)      & 1.415\,429\ldots&            &               &          & 16.595\,404\ldots&\\
\hline
1024 (T)     & 1.414\,89(4)  & 0.194\,9(15) & 0.919(15)     & 0.240(10)& 18.28(8) & 1.163(6)     & 148.8(19) \\
1024 (C)     & 1.414\,826(6) & 0.193\,07(12)& 0.904\,6(11)  &          & 18.35(3) & 1.168\,1(4)  & 146.1(4)\\
1024 (E)     & 1.414\,821\ldots&            &               &          & 18.361\,348\ldots       \\
\hline 
\end{tabular*}
\end{table}

\subsection{Canonical averages at the critical temperature and zero field}\label{results-Tc}
We present in Table~\ref{tab:results-Tc} our values for the canonical averages 
of several physical quantities. We have compared with the exact
results for $u_\beta$ and $C$ in finite lattices computed with the expressions
given by Ferdinand and Fisher~\cite{FERDINAND-FISHER}.
We have also run simulations with a Swendsen-Wang cluster algorithm (we use our own 
implementation, based on the one distributed with~\cite{AMIT-VICTOR},
but our results are compatible with those of~\cite{SALAS-SOKAL}).

From Table~\ref{tab:results-Tc} we can confirm that TMC is capable
of producing very accurate results. The relative errors for $\chi_2$ and $B$ 
scale as $L$. This can be accounted for by noticing that both 
are completely determined by $p(\hat m)$, recall Section~\ref{momentos}.
Indeed, while $\hatah$ is self averaging and virtually
free of critical slowing down (meaning that for a fixed simulation length
its error scales as $1/\sqrt{N}$),
we are multiplying $\varOmega_N(\hat m,\beta)$ by a factor of $N$, yielding
an overall $\sqrt{N}$ scaling for the errors.

As we said in the Introduction, TMC is 
not meant to be a competitor to cluster algorithms for the Ising model
without magnetic field. For example, the CPU time to compute 
each of the $77$ simulations at fixed $\hat m$ for $L=1024$ is
about a third of what we needed for the whole Swendsen-Wang simulation, which
is also more precise. 

\subsection{Finite Size Scaling and the peaks of  $p_1(m)$}\label{peaks}
Let us obtain an accurate estimate of the critical exponent ratio $\beta/\nu$
(known to be $1/8$) in a way that would not be competitive
for a canonical computation.

Our starting point is the Finite Size Scaling formula for
an arbitrary observable $\aO$ (see, e.g.,~\cite{AMIT-VICTOR}),
as we get closer to the critical point ($\beta=\beta_\mathrm{c}$, $h=0$):
\begin{equation}\label{eq:FSS}
\medio{\aO}_t(h) = L^{-x_\aO/\nu} \left[ f_\aO( L^{1/\nu} t, L^{y_h} h)+ \ldots\right],\qquad t =  \frac{\beta_\mathrm{c}-\beta}{\beta_\mathrm{c}}.
\end{equation}
Here the dots represent possible corrections to scaling. We shall work at  $h=0$, 
so we only have to consider the first argument of the scaling function $f_\aO$.

Elaborating equation \eqref{eq:FSS} and recalling that $\beta$
is the critical exponent for the magnetisation, it can be shown that~(see, e.g., \cite{AMIT-VICTOR})
\begin{equation}
\tilde p_1(m,\beta_\mathrm{c};L) = L^{\beta/\nu} \tilde f(L^{\beta/\nu} m),
\end{equation}
where $\tilde p_1(m,\beta_\mathrm{c};L)$ is some smoothed
version of $p_1(m,\beta_\mathrm{c};L)$, equation~\eqref{eq:p1}.
Recalling that $p(\hat m,\beta_\mathrm{c};L)$, equation~\eqref{eq:p}, is just one of such smoothings,
we can substitute this expression by
\begin{equation}
p(\hat m,\beta_\mathrm{c};L) = L^{\beta/\nu} f\bigl(L^{\beta/\nu} (\hat m-1/2)\bigr).
\end{equation}
In particular, the pdf has two maxima at $m_\text{peak}^\pm+\tfrac12$, whose
scaling behaviour is expected to be $m^\pm_\text{peak}\propto L^{-\beta/\nu}$.

Recall that we measure directly $\hatah$, which
is the derivative of the logarithm of $p(\hat m)$ and thus is exactly zero 
at these peaks. Therefore, for each of the maxima we just have to identify the two
consecutive points of the grid such that $\medio{\hatah}_{\hat m_i,\beta} >0$ 
and $\medio{\hatah}_{\hat m_{i+1},\beta} <0$ and find the root of the cubic spline
joining them (we have also used jackknife blocks to estimate the errors).  

The position of these peaks is a canonical observable, but one that
is more easily obtained through the tethered ensemble (in a canonical simulation
we would have had to construct $p_1(m)$ and locate its maximum directly,  
a harder problem than finding a zero).

The results of this analysis for the simulations described in Section~\ref{parameters}
are collected in Table~\ref{tab:pico}. Notice that 
even with our relatively coarse $\hat m$ sampling we have been able 
to determine the position of the peaks with a precision ranging from $0.013~\%$ 
to $0.46~\%$. If we had wanted to give a very precise value for these points, 
we could have done so with a rather small investment in CPU time, as we only need 
precise simulations at two values of $\hat m$, conveniently chosen.

By fitting $m_\text{peak}^{\pm}(L)$ to a power law we can obtain 
an estimate of $\beta/\nu$. Table~\ref{tab:beta-nu} 
gathers the results of fitting the data of Table~\ref{tab:pico}
to a power law, for lattice sizes $L\geq L_\text{min}$. We see 
that for $L_\text{min}=32$ the power law does not represent
the curve adequately. To give an error estimate 
that represents both systematical and statistical sources we follow the criterion
explained in~\cite{UCM-beta}: when two consecutive values
are first compatible, we give the most precise as central value,
but with the bigger error.
For the negative magnetisation peak we find $\beta/\nu = 0.123\,9(11)$
and for the positive peak $\beta/\nu=0.124\,5(10)$.

Following~\cite{SALAS-SOKAL}, we can even go further and try to characterise the corrections
to scaling. We assume that in the Ising model
in $D=2$ the dominant corrections to scaling are analytical
\begin{equation}
m_\text{peak}^{\pm} = L^{-\beta/\nu}[ A^\pm + B^\pm L^{-\varDelta}],\qquad \varDelta = 7/4.
\end{equation}
We have fitted our points for all lattices to this expression, fixing 
the exponents to their exact values and varying $A^\pm$ and $B^\pm$.
We have obtained $\chi^2/\text{d.o.f.} = 0.9858/4$ for the negative 
peak and $\chi^2/\text{d.o.f.} = 2.825/4$ for the positive one.

\begin{table}[h]
\centering
\caption{Position of the peaks of the probability density 
function of the magnetisation for several lattice sizes. 
As we are at the critical point, the values in the table
extrapolate to zero when $L\to\infty$ (compare with Section~\ref{results-ferromagnetic}, 
where we study the ferromagnetic region).}
\label{tab:pico}
\smallskip

\begin{tabular*}{.7\columnwidth}{@{\extracolsep{\fill}}rll}
\hline
\boldmath $L$ & \multicolumn{1}{c}{\boldmath $-m^{-}_\mathrm{peak}$} &
 \multicolumn{1}{c}{\boldmath $m_\mathrm{peak}^{+}$}\\
\hline
32  & 0.764\,01(10) & 0.764\,31(11) \\
64  & 0.702\,86(18) & 0.703\,0(2)     \\
128 & 0.645\,3(3)   & 0.645\,1(4)   \\
256 & 0.592\,1(7)   & 0.591\,0(7)   \\
512 & 0.541\,9(12)  & 0.542\,7(9)   \\
1024& 0.499(2)      & 0.500(2)      \\
\hline
\end{tabular*}
\end{table}

\begin{table}[h]
\centering
\caption{Fits of $m_\text{peak}^{\pm}(L)$ to a power law, 
$m_\text{peak}^\pm = A L^{-x}$,
including all lattice sizes $L\geq L_\text{min}$. We give
the computed exponent and the chi square per degree of freedom. As discussed 
in the text, for small lattices we detect corrections
to scaling. Our fits converge to the exact
result, $\beta/\nu=0.125$.}\label{tab:beta-nu}
\smallskip

\begin{tabular*}{.7\columnwidth}{@{\extracolsep{\fill}}rlclc}
\cline{2-5}
& \multicolumn{2}{c}{\boldmath $m^{-}_\mathrm{peak}$} &  \multicolumn{2}{c}{\boldmath $m_\mathrm{peak}^+$} \\
\hline
\boldmath $L_\mathrm{min}$ & 
\multicolumn{1}{c}{\boldmath $\beta/\nu$}
 & \boldmath $\chi^2/\mathrm{d.o.f.}$
 &\multicolumn{1}{c}{\boldmath $\beta/\nu$} & \boldmath $\chi^2/\mathrm{d.o.f.}$  \\
\hline
32  & 0.121\,7(3)   &  23.44/4  &  0.122\,4(3)       &  27.85/4  \\ 
64  & 0.123\,9(5)   &  2.027/3  &  0.124\,5(5)       &  2.087/3   \\
128 & 0.125\,0(11)  &  0.7569/2 &  0.124\,6(10)      &  2.053/2   \\
256 & 0.126(4)      &  0.6456/1 &  0.122\,0(23)      &  0.3248/1      \\
\hline
\end{tabular*}
\end{table}

\section{The scaling paramagnetic region: the renormalised coupling constants}\label{results-paramagnetic}
We shall show here that TMC works as well in the paramagnetic phase
in the scaling region. For definiteness, we shall study the renormalised
coupling constants (see~\cite{ZINN-JUSTIN,PELISSETTO-VICARI}). These
constants are notoriously difficult to compute using Monte Carlo methods, 
so they provide a good challenge for our formalism.

Let us consider the $D=2$ Ising model in its thermodynamical limit.
Then we can define the Gibbs free energy by
\begin{equation}
G(t,h) = \lim_{N\to\infty} \frac1N \log[ Z(t,h)],\qquad t = \frac{\beta_\mathrm{c} - \beta}{\beta_\mathrm{c}}
\end{equation}
and its Legendre transform, the Helmholtz free energy ($m_t=\medio{\am}_t$),
\begin{equation}
F (t,m_t) = \max_h \bigl[ m_t h - G(t,h)\bigr].
\end{equation}
From a field theoretical point of view, the latter can be expanded 
as a series in the renormalised coupling constants
$g_{2n}$ (here we shall use the definitions of~\cite{CASELLE-etal})
\begin{align}
F(t,m_t) - F(t,0)
        &= \frac{1}{\xi^2} \biggl( \frac12 \phi^2+\sum_{n=2}^\infty
                                   \frac{g_{2n}}{(2n)!}\phi^{2n}\biggr)\\
        &= - \frac{\chi_2^2}{\chi_4}\left(\frac12 z^2 + \frac1{4!} z^4
           + \sum_{n=3}^\infty \frac{r_{2n}}{(2n)!} z^{2n} \right).
\end{align}
In these equations,
\begin{equation}
\phi^2 = \frac{\xi(t)^2}{\chi_2(t)}  m_t^2,\qquad \qquad
z^2    = - \frac{\chi_4(t)}{\chi_2(t)^3} m_t^2.
\end{equation}
We would like to express the $g_{2n}$ in a form suitable 
for our lattice simulations. To do this we remember definition~\eqref{eq:chi-2n}, 
according to which $G(t,h)$ can be immediately 
represented as a Taylor expansion with coefficients $\chi_{2n}$,
\begin{equation}\label{eq:logZ}
G(t,h)-G(t,0) = \sum_{n=1}^\infty \frac{\chi_{2n}}{(2n)!} h^{2n}.
\end{equation}
Combining all these equations we arrive at the following explicit formulas:
\begin{equation}
g_4    = - \lim_{t\to0^+} \frac{\chi_4(t)}{\chi_2(t)^2 \xi(t)^2}, \qquad\qquad
g_{2n} = r_{2n} (g_4)^{n-1},
\end{equation}
with
\begin{align}
r_6    &= 10 - \lim_{t\to0^+} \frac{\chi_6(t) \chi_2(t)}{\chi_4(t)^2},\\
r_8    &= 280 + \lim_{t\to0^+} \left[\frac{\chi_8(t) \chi_2(t)^2}{\chi_4(t)^3} - 56 \frac{\chi_6(t) \chi_2(t)}{\chi_4(t)^2}\right]
\end{align}
There has been a great deal of interest about these coupling constants
and many precise computations have been performed, both with
field theoretical and numerical methods. Balog et al.~\cite{BALOG-etal}
arrive at $g_4 = 14.697\,5(1)$ with a form factor approach while 
Caselle et al.~\cite{CASELLE-etal} give $g_4=14.697\,323(20)$,
$r_6 = 3.678\,66(3)(2)$ and $r_8=26.041(8)(3)$ using transfer matrix techniques.
There are fewer Monte Carlo determinations of these quantities~\cite{BALOG-etal,KIM}.

In this Section we shall reproduce the Monte Carlo computations of~\cite{KIM}
with the TMC method to give our own estimate of the first coupling constants.
A Monte Carlo determination of the $g_{2n}$ can obviously not be 
directly computed for an infinite lattice. Instead, we have to perform
a double limit $\lim_{t\to0^+}\lim_{L\to\infty}$, that is, run simulations
very close to, but above, the critical temperature at increasing 
values of~$L$ until the results are stable (within our errors). 
Ref.~\cite{KIM} concludes that $L=100$ is a big enough lattice. 
In order to compare our results to those of that reference, we have worked
at the same temperature, $\beta=0.42$.

We have used $191$ uniformly spaced $\hat m$ in $[-0.45,1.45]$, with $10^7$ Monte
Carlo sweeps each. Table~\ref{tab:gn} summarises our results, which are compatible
with those of~\cite{KIM} but more precise. These calculations involve 
working with high moments of the magnetisation (we have used the indirect
method explained in Section~\ref{momentos}, combining the cumulants 
of $\ar$ and $\hatam$). The table also compares the tethered results 
with a Swendsen-Wang simulation (SW). Notice that in the latter 
the errors increase very quickly when we consider high powers
of $\am$. For instance, for $L=100$ the error in the susceptibility 
is almost the same for the TMC and the SW methods, while 
for $g_8$ the error of the latter is almost ten times higher.
Our whole TMC simulation for $L=100$ required about 100 hours of computer
time, while the Swendsen-Wang one was completed in about 10.
If we use `improved' or cluster estimators (SWC)~\cite{CLUSTER-ESTIMATORS},
however, the advantage of the tethered algorithm disappears. 

\begin{table}[h]
\centering
\caption{The renormalised coupling constants. TMC: Tethered Monte Carlo;
SW: Swendsen-Wang;  SWC: Swendsen-Wang with cluster estimators.
For $L=100$ we also give the results of Ref.~\cite{KIM}, computed 
with a single cluster Monte Carlo method.} \label{tab:gn}
\smallskip 

\begin{tabular*}{\columnwidth}{@{\extracolsep{\fill}}ccccccccc}
\cline{2-9}
& \multicolumn{3}{c}{\boldmath $L = 50$}&  & \multicolumn{4}{c}{\boldmath $L=100$}\\
&\bfseries TMC & \bfseries SW & \bfseries SWC& & \bfseries TMC & \bfseries SW & \bfseries SWC  & \bfseries Ref. \cite{KIM} \\
\hline
\boldmath $-\medio{\au}_\beta$     & 1.228\,238(14) & \multicolumn{2}{c}{1.228\,31(4)}  & & 1.226\,067(7)&  \multicolumn{2}{c}{1.226\,076(8)} &  \\     
\boldmath $\chi_2$ & 196.85(9)      &  196.96(15)  & 196.91(5)    & & 203.78(11)   &  204.07(10)    & 203.92(2)     & 204.4(3)  \\        
\boldmath $\xi_1$  & 11.749(5)      &  11.756(9)   & 11.753(3)    & & 11.888(11)   &  11.907(10)    & 11.893\,2(10) & 11.90(2)  \\
\boldmath $r_6$    & 4.446\,2(9)    &  4.449(4)    & 4.446\,9(10) & & 3.70(6)      &  3.73(8)       & 3.731(6)      & \\   
\boldmath $r_8$    & 39.76(3)       &  39.83(13)   & 39.77(3)     & & 26.2(6)      &  24(3)         & 26.47(18)     & \\
\boldmath $g_4$    & 12.817(6)      &  12.80(3)    & 12.812(7)    & & 14.66(5)     &  14.69(9)      & 14.673(8)     & 14.60(16)\\
\boldmath $g_6$    & 730.4(5)       &  729(3)      & 729.9(9)     & & 794(9)       &  806(24)       & 803.3(16)     & $8.5(4)\cdot10^2$ \\
\boldmath $g_8/10^{4}$ & 8.370(8)   &  8.36(6)     & 8.363(16)    & & 8.25(13)     &  7.5(11)       & 8.34(7)       & $8.8(10)$\\
\hline
\end{tabular*}
\end{table}

\section{Results at \boldmath$\beta_\mathrm{c}$, $h\neq 0$}\label{results-h}
An interesting feature of TMC is that it allows us to obtain
accurate data in the presence of a magnetic field. In order to do this, we reanalise the data from the simulations
described in Section~\ref{parameters}. Let us stress the fact that we do not have to
run any new simulations at all, we simply use the tethered values computed 
at zero magnetic field and modify the effective potential as in equations~\eqref{eq:Omega-h}
and~\eqref{eq:valor-canonico-h}.

We will typically be interested not in computing the equivalent
of Table~\ref{tab:results-Tc} for a particular value of~$h$, but
in drawing a curve $\langle \aO\rangle_{\beta}(h)$. When doing 
this, we can improve our statistical errors somewhat if we take into
account that the observables can be either odd or even in~$h$. 
This means that we can (anti)symmetrise the curves:
\begin{align}
\langle\aO\rangle^\mathrm{odd}_\beta(h) &= \frac{\medio{\aO}_\beta(h) - \medio{\aO}_\beta(-h)}{2}\ ,\label{eq:odd}\\  
\langle\aO\rangle^\mathrm{even}_\beta(h) &= \frac{\medio{\aO}_\beta(h) + \medio{\aO}_\beta(-h)}{2}\ .\label{eq:even}
\end{align}
Now, if the data were completely uncorrelated, so that $\medio{\aO}_\beta(\pm h)$ were 
independent, this operation would imply a $1/\sqrt2$ reduction in the statistical error.
Actually, this is not the case: the individual values for $\pm h$ are very strongly
correlated and the symmetrisation reduces only very slightly the error for even quantities.
For odd observables, however, the fact that we are computing a difference rather than a sum
yields a very significant decrease in the statistical error (around a factor of~$10$, specially
for small values of $h$). We shall use equations~\eqref{eq:odd} and~\eqref{eq:even}, but
we shall drop the explicit `odd' or `even' superscripts.

\begin{figure}[t]  
\centering
\includegraphics[height=.7\linewidth,angle=270]{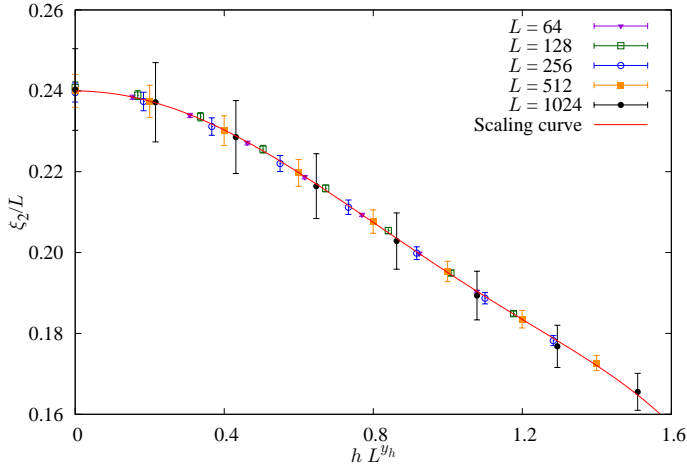}
\caption{Finite Size Scaling with $\xi_2(h)$. We represent the
scaling function by a fit to $f_\xi(x) = a_0 + a_2 x^2 +a_4 x^4 + a_6 x^6$.
The value of the diagonal chi square per degree of freedom is $\chi^2_\mathrm{d}/\text{d.o.f} = 3.978/36$,
which confirms that the points fall on a smooth curve within their errors.}
\label{fig:scaling-xiL}
\end{figure}

The first observable we shall consider is the correlation length~$\xi_2(h)$, defined
in equation~\eqref{eq:xi2}. This is even in~$h$, so we can use the symmetrised 
version of equation~\eqref{eq:even}. In this case, unlike previous sections, 
no data for~$\xi_2(h)$ are readily available in a finite lattice and doing our own
cluster simulations would be harder (as we pointed out in the Introduction, 
the currently most popular methods in the $h\neq0$ regime are transfer matrix techniques).
Nevertheless, we can provide a very good check of consistency using the FSS
equation~\eqref{eq:FSS}. Indeed, given
that our simulations have been carried out at the critical temperature, we 
can consider the critical behaviour as $h\to0$.  In our case, we are working at precisely
$\beta=\beta_\mathrm{c}$, so the first variable in the scaling function disappears and we have 
($x_\aO=\nu$ for the correlation length) 
\begin{equation}\label{eq:FSS2}
\xi_2/L \simeq f_{\xi_2}(L^{y_h} h).
\end{equation}
The critical exponent $y_h$ is $15/8$ for the two dimensional Ising model 
and the function $f_{\xi_2}$ is expected to be very smooth. 
As we can see in Fig.~\ref{fig:scaling-xiL}, equation~\eqref{eq:FSS2} is 
perfectly valid in our case, once we discard the data for $L\leq 32$.
 At a first glance, it may seem that we
have even overestimated our errors for $L=1024$, but remember
that the points are very strongly correlated. The universal scaling curve
is well represented by a sixth order even polynomial, with a value
of diagonal $\chi^2_\mathrm{d}/\text{d.o.f.}$ of $3.978/36$.\footnote{%
The points of the graph are very strongly correlated, so 
their covariance matrix is singular and the $\chi^2$ parameter
of the fit cannot be computed. Instead, we restrict
ourselves to the diagonal elements of the covariance
matrix and minimise the resulting `diagonal' $\chi^2_\mathrm{d}$. 
Usually, the degrees of freedom are computed as 
number of points minus the number of parameters. 
In our case this does not correspond
to the actual degrees of freedom of the fit, but 
we have maintained the usual notation.}

Thus, the universal scaling function $f_{\xi_2}(x)$ for $x\lesssim1.5$  is extremely well represented by\footnote{%
To estimate the errors we have computed a fit for each jackknife block.}
\begin{equation}
f_{\xi_2}(x) = a_0 + a_2 x^2 + a_4x^4+ a_6x^6,
\end{equation}
with
\begin{align}
a_0 &= 0.239\,9(2),& a_2&=-0.063\,9(4), & a_4&=0.023\,5(7), & a_6 &= -0.004\,5(4).
\end{align}

We now consider the magnetisation, $m_{\beta_\mathrm{c}}(h)=\langle \am\rangle_{\beta_\mathrm{c}}(h)$. Contrary
to the correlation length, the magnetisation is an odd 
function of~$h$. This means that by antisymmetrising our results for $\pm h$ 
as in equation~\eqref{eq:odd} we can greatly reduce the error. In this case
we again lack a convenient way of generating the same curve with a different method, 
but we still can check the validity of our results. To do this, we differentiate
equation~\eqref{eq:logZ} and notice that the coefficients of the Taylor expansion of $\langle m\rangle_{\beta_\mathrm{c}}(h)$
are none other than the $2n$ point correlation functions at zero momentum:
\begin{equation}\label{eq:desarrollo-m-h}
m_{\beta_\mathrm{c}}(h)=  \medio{\am}_{\beta_\mathrm{c}}(h)=
\frac1N \frac{\partial \log Z}{\partial h} =\chi_2 h +
\frac{\chi_4}{3!} h^3 + \frac{\chi_6}{5!} h^5 + \frac{\chi_8}{7!}h^7 + \ldots .
\end{equation}
This expansion provides a new way of computing~$\chi_{2n}$: we just generate a reasonable
number of points of the $m(h)$ curve, which we parameterise with a 
truncated version of equation~\eqref{eq:desarrollo-m-h}. The choice of 
values for $h$ is somewhat delicate: if we use very small 
magnetic fields we will only be able to appreciate the first few coefficients
but if we go too far in $h$ we would need to have sampled the tails of 
the pdf of $\hat{m}$ very precisely. We have found that magnetic fields 
up to $\sim (\chi_2)^{-1}$ provide a good compromise. 

A second difficulty is that
posed by the correlations among the points of the curve. These are so strong
that the covariance matrix turns out singular, which bars us from
obtaining a fit and its errors by minimising $\chi^2$. Instead, we have 
computed an odd interpolating polynomial with a finite difference formula
(this gives us as many $\chi_{2n}$ as points) and we have tried to control
the correlation by estimating the errors with the jackknife method. This takes
care of the statistical error. To control the systematic error, we have 
both reduced the range in $h$ and varied the number of points to check
whether the parameters were stable. We have found that if we use this
method with $n$ points the last one or two coefficients are 
usually unreliable (i.e., their value changes beyond the error 
bars if we add another point). This only means that
if we want to obtain $n$ physically meaningful parameters,
we should compute at least $n+2$ points. 
We want to compute the nonlinear susceptibilities up to $\chi_{8}$, so to be safe we have 
used $7$ points for each lattice size, equally spaced at intervals of $(10\chi_2)^{-1}$,
where $\chi_2$ is the susceptibility computed from the simulation at $h=0$.

Table~\ref{tab:fit-mh} summarises our results for the $\chi_{2n}$, computed
from the measurements at $h=0$ and from the finite difference formula at $h\neq0$.
We see that both series of values are compatible, but that the former
are more precise.

\begin{table}
\centering
\caption{Nonlinear susceptibilities from direct measurements at $h=0$ (M) and
from a finite difference formula for $m_{\beta_\mathrm{c}}(h)$ as
a function of the magnetic field (F).}
\label{tab:fit-mh}
\smallskip

\begin{tabular*}{\columnwidth}{@{\extracolsep{\fill}}rllll}
\hline
\multicolumn{1}{c}{\boldmath $L$} & 
\multicolumn{1}{c}{\boldmath $N^{-1}\chi_2$} &
\multicolumn{1}{c}{\boldmath $N^{-2}\chi_4$} &
\multicolumn{1}{c}{\boldmath $N^{-3}\chi_6$} &
\multicolumn{1}{c}{\boldmath $N^{-4}\chi_8$}\\
\hline
16 (M)        & 0.545\,43(6)       & $-0.545\,72(13)$     & 2.265\,7(8)        & $-20.059(10)$ \\ 
16 \;(F)      & 0.545\,43(6)       & $-0.545\,7(2)$       & 2.262\,8(19)       & $-19.70(14)$   \\
\hline
32 (M)        & 0.459\,00(10)      & $-0.386\,1(2)$       & 1.348\,5(10)       & $-10.042(10)$\\
32 \;(F)      & 0.459\,00(10)      & $-0.386\,2(3)$       & 1.348\,4(18)       & $-10.00(2)$  \\
\hline
64 (M)        & 0.386\,19(18)      & $-0.273\,3(3)$       & 0.803\,1(13)       & $-5.032(11)$ \\ 
64 \;(F)      & 0.386\,19(18)        & $-0.273\,8(5)$       & 0.805(2)           & $-5.02(2)$   \\ 
\hline
128 (M)       & 0.324\,4(3)        & $-0.192\,8(4)$       & 0.475\,8(17)           & $-2.504(12)$ \\
128 \;(F)     & 0.324\,4(3)        & $-0.194\,3(7)$       & 0.481(3)           & $-2.52(2)$   \\ 
\hline
256 (M)       & 0.272\,8(6)        & $-0.136\,2(7)  $     & 0.283(2)           & $-1.250(12)$ \\
256 \;(F)     & 0.272\,5(6)        & $-0.135\,8(15) $     & 0.280(6)           &$ -1.20(4)  $ \\
\hline                                                                                     
512 (M)       & 0.229\,3(7)        & $-0.096\,4(7)  $     & 0.168(2)           &$  -0.625(9)$   \\
512 \;(F)     & 0.229\,3(7)        & $-0.096\,0(12) $     & 0.166(4)           &$ -0.60(2)  $ \\
\hline                                                                                     
1024 (M)      & 0.194\,9(15)       & $-0.069\,8(13) $     & 0.104(3)           &$  -0.328(12)$  \\
1024 \;(F)    & 0.194(3)           & $-0.063(6)     $     & 0.08(2)            &$ -0.2(3)   $ \\
\hline                                                                                     
\end{tabular*}                       
\end{table}                          

\begin{figure}[t]
\includegraphics[height=\linewidth,angle=270]{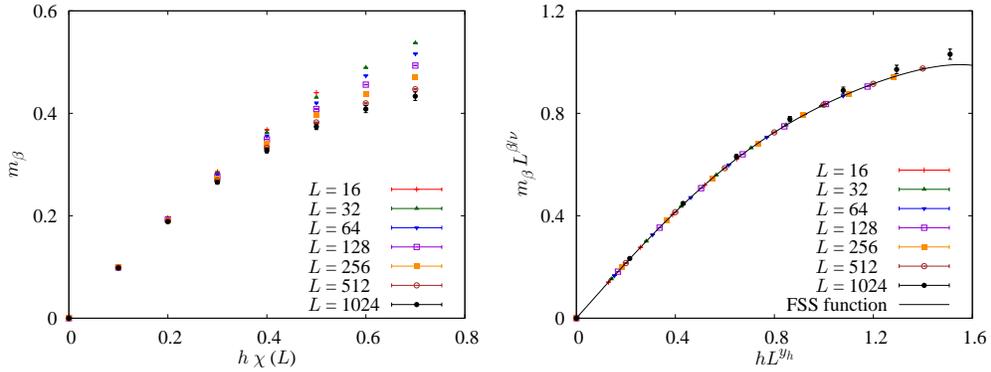}
\caption{Magnetisation at nonzero magnetic field for different
lattice sizes. We provide two representations. The one on the left
is plotted in terms of the applied field times the susceptibility
at $h=0$. This way we can appreciate the linear and nonlinear regimes.
The graph on the right is the FSS curve for $m_\beta(h)$ (a fit
to a seventh order odd polynomial, with diagonal $\chi^2_\mathrm{d}/\text{d.o.f}=41.85/31$).}
\label{fig:mh}
\end{figure}

We can also do a FSS analysis with the magnetisation. Notice that at small applied 
fields $m_\beta(h)\simeq \chi_2 h$.
If we increase the field, however, we can appreciate a deviation 
from the linear behaviour (see the left graph of Fig.~\ref{fig:mh}). If we want to 
collapse all the curves on one we must take the critical exponent 
of the magnetisation, $-\beta$, into account:
\begin{equation}\label{eq:FSS-m}
m_{\beta_\mathrm{c}}(h) \simeq L^{-\beta/\nu} f_m(L^{y_h} h).
\end{equation}
The function $f_m$ is depicted on the right panel of Fig.~\ref{fig:mh}. 
We fit it to an odd polynomial $f_m(x) = a_1 x +a_3x^3 + a_5x^5+ a_7x^7$
for lattices $L\geq 64$, as we did for the correlation length. The value of
the diagonal $\chi^2_\mathrm{d}/\text{d.o.f.}$ for this fit is $41.85/31$. The last point
of the curve, which corresponds to the highest magnetic field for $L=1024$,
seems to show a small deviation from the curve. The reason may be that 
we are taking the range of the scaling variable too far and that scaling 
corrections start to act. Remember that we had spaced the values of $h$ in
units of $(10\chi_2)^{-1}$, so that the representation on 
the left scaled properly, which is not be best choice 
for the FSS analysis.

\begin{figure}[t]
\centering
\includegraphics[height=.7\linewidth,angle=270]{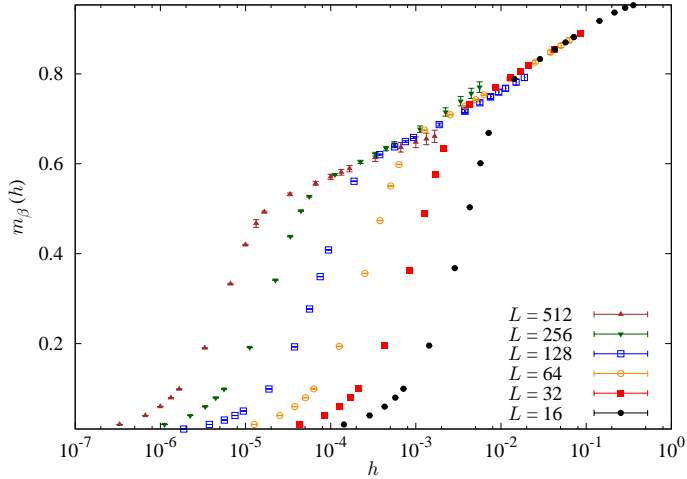}
\caption{Magnetisation at the critical point for several lattices
and an $L$ independent range of magnetic field values
(notice the logarithmic scale of the horizontal axis).
}\label{fig:h-grande}
\end{figure}

It is interesting to examine the behaviour of the magnetisation
with small but $L$ independent magnetic fields (Fig.~\ref{fig:h-grande}).
We observe two well differentiated scales: an FSS regime, 
where the slope of the curve is very large, and a thermodynamic limit, 
where the curves merge. The range of $h$ is limited
by the largest obtained $\medio{\hatah}_{\hat m,\beta}$ in 
the simulated $\hat m$ grid (this is the reason for the rapid
growth of the error bars for the final points, specially 
noticeable in the largest lattices).
Due to the small value of $\beta/\nu$, 
the displacement of the curves for small $h$ is almost 
linear in $\log L$, see equation~\eqref{eq:FSS-m}. 

We can repeat the process described here for other observables, such as
the energy or specific heat. However, our computation can be inefficient
if we do not prepare it carefully. Remember that for $h=0$ the 
canonical average was dominated by two peaks, whose position was
determined by the value of $\hat m$ such that $\medio{\hatah}_{\hat m} = 0$.
Now the value with a nonzero magnetic field will be even more dependent
on a saddle point, determined by
\begin{equation}
\medio{\hatah}_{\hat m} = h.
\end{equation}
In practice,
this means that the way to obtain a precise result for, say, the energy
at a given field $h$ is to first estimate the value of $\hat m$ such that
the previous equation is satisfied and then run a long simulation there.

\section{The broken symmetry phase}\label{results-ferromagnetic}
Let us now consider the $\beta>\beta_\mathrm{c}$ regime. In this situation,
the infinite system shows a nonzero expectation value
for the order parameter, $m_{\beta>\beta_\mathrm{c}} = \medio{\am}_{\beta>\beta_\mathrm{c}}\neq 0$,
even in the absence of an external magnetic field. This may seem
incompatible with the partition function~\eqref{eq:Z}, where
the configurations $\{\sigma_{\boldsymbol x}\}$ and
$\{-\sigma_{\boldsymbol x}\}$ occur with equal probability. 
The well known solution for this apparent paradox 
is spontaneous symmetry breaking~\cite{ZINN-JUSTIN}, whose mathematical formulation
involves considering a small magnetic field (which establishes 
a preferred direction) and taking the double limit
\begin{equation}\label{eq:symmetry-breaking}
\medio{\am}_{\beta,\infty} = \lim_{h\to 0} \lim_{L\to\infty}\medio{\am}_{\beta,L}.
\end{equation}
The order of the two limits is crucial: were we to reverse it the magnetisation
would always vanish. We see then that the symmetry of our model 
complicates the definition of a broken symmetry phase for finite lattices
in the canonical ensemble. The traditional workaround consists in considering
not the magnetisation $\am$, but its absolute value $|\am|$. 

The tethered ensemble provides a cleaner concept of broken symmetry phase.
Consider the pdf of $\hat m$, as in Fig.~\ref{fig:tethered}.  In the ferromagnetic
phase the corresponding graph will again have two peaks, but now these will
be much narrower and higher, approaching two Dirac deltas in the thermodynamic
limit. Suppose we want to perform the double limit of
equation~\eqref{eq:symmetry-breaking}. This would involve introducing a small
magnetic field which would shift the origin of $\hatah$~\eqref{eq:canonical-average-h}.
The neighbourhood of one of the peaks would then become exponentially
suppressed and eventually disappear in the thermodynamic limit. Thus, we can mimic
the effect of equation~\eqref{eq:symmetry-breaking} by considering 
only one of the two peaks from the beginning. This would not work
below $\beta_{\mathrm{c}}$, as there the peaks extrapolate to $m=0$ (recall
Section~\ref{peaks}). In a more complex model,
 the $L$ evolution of the right peak should be checked, 
in order not to mistake a paramagnetic
phase for a ferromagnetic one.
This criterion suggests that TMC could be a powerful method
for determining the order of magnetic phase transitions.

This procedure has the 
considerable advantage that it works for any lattice size. In this section we 
have chosen the peak of positive magnetisation. We will illustrate
it by considering the thermodynamic limit in the canonical
ensemble in Section~\ref{thermodynamic-limit} and 
by studying the equivalence between the tethered and canonical
ensembles in Section~\ref{ensemble-equivalence}.

\subsection{The thermodynamic limit}\label{thermodynamic-limit}
We have run simulations for lattice sizes $L=128,256,512,1024$ 
at $\beta=0.4473>\beta_{\mathrm c}$. This temperature was chosen because
we estimated that it would roughly mirror the value
of the correlation length\footnote{In the ferromagnetic regime
$\xi_1$ is not a good definition and we always use $\xi_2$, see equation~\eqref{eq:xi2}.} for our paramagnetic simulations. 

Following the previous discussion, we have worked in the
$\hat m > 0.5$ (positive magnetisation) region, where there is only one peak.
An appropriate sampling of $\hat m$ is even more important in this phase (but
easier to optimise) than in the situation described in detail in Section~\ref{sampling}.
The reason is that the peak is now so narrow that a choice of $\hat m$ spaced
as in the aforementioned section would not only be completely wasteful,
but may also completely fail to sample the peak.

In the case of the Ising model,
we know Yang's exact solution $m_\textsc{y}(\beta)$ for the magnetisation of the
infinite system~\cite{YANG}. The positive peak for $p(\hat m)$ will then be very close
to $m_\textsc{y}(\beta)+\tfrac12$ and get closer as we increase~$L$. With this
information in hand, we can adequately reconstruct the effective potential
by running simulations in a small neighbourhood of $m_\textsc{y}(\beta)+\tfrac12$.
For a different model, where we would lack the knowledge of the peak's position
in the thermodynamic limit, we can just run simulations with a very fine 
grid for some small and essentially costless lattice size and infer from them
an efficient distribution of points for the larger systems. 
\begin{figure}[t]
\centering
\includegraphics[height=.7\linewidth,angle=270]{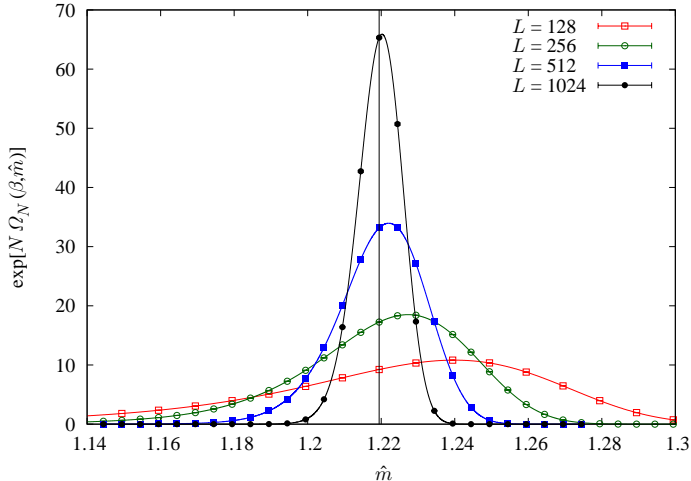}
\caption{Computation at $\beta=0.4473>\beta_{\mathrm{c}}$ (ferromagnetic phase).
The peak of the pdf of $\hatam$ gets narrower and closer
to $m_\textsc{y}+1/2$ as $L$ increases (the vertical line, where
$m_\textsc{y}$ is Yang's magnetisation for the infinite system
at $\beta=0.4473$~\cite{YANG}).
Compare the scale of the $x$ axis and the height of the peaks
with those of Fig.~\ref{fig:tethered}.}
\label{fig:Yang}
\end{figure}

We have represented $p(\hat m)$ for all the simulated lattices in Fig.~\ref{fig:Yang}, which
includes the whole range of $\hat m$ for $L\geq512$ (for the smaller lattices we have used
a somewhat larger interval). It is interesting to compare the scale on the $x$ axis with
that of Fig.~\ref{fig:tethered}. As will be discussed in detail in
Section~\ref{ensemble-equivalence}, the peak
approaches $m_\textsc{y}(\beta)+ \tfrac12$ (the vertical line) as $L$ increases. Table~\ref{tab:results-BSP}
compares the values of the energy and specific heat obtained in our simulations
with the exact values given in~\cite{FERDINAND-FISHER}. Notice how very small simulated
ranges of $\hat m$ ($\Delta \hat m = \hat m_\text{max} - \hat m_\text{min}$) yield very accurate results. 
In fact, we can see that for the $L=1024$ lattice we obtain a more precise determination
for the energy with $27$ points than what we obtained at the critical temperature
with $77$ (we still perform $10^7$ Monte Carlo sweeps in each point). This result is even more
impressive if we consider that some of these $27$ points, being deeply inside the tails of the distribution,
do not have any effect whatsoever 
in the average with our error (of course, we do not know this until we have run the simulation 
and seen the actual width of the peak).

From Table~\ref{tab:results-BSP} we can conclude that the thermodynamic limit has already been
reached for $L=256$, at least to the level indicated by our errors.
Our whole computation for $L=512$ required about $270$ hours of computer time.
For comparison, a $30$ hour long Swendsen-Wang computation 
of the correlation length for $L=512$ gives $\xi_2=11.8(2)$. We see that
the ratio of computation time for both methods has changed significantly 
from the critical point, where the advantage of the cluster algorithm was
much greater.

\begin{figure}[t]
\centering
\includegraphics[height=.7\linewidth,angle=270]{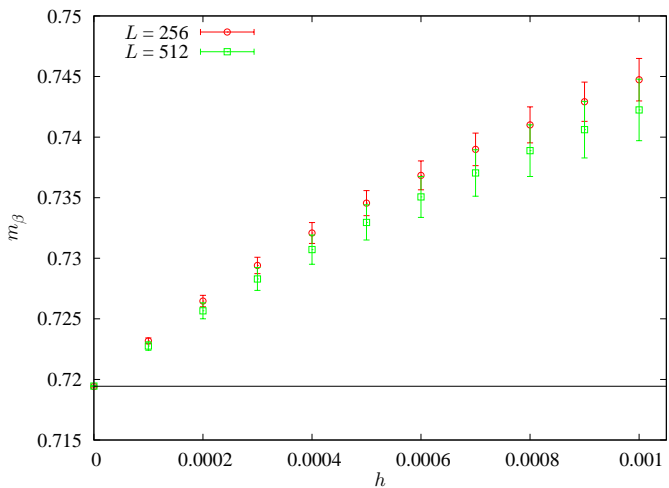}
\caption{Magnetisation in the presence of a magnetic field in the ferromagnetic
regime for $\beta = 0.4473$. The horizontal line is the exactly known
thermodynamic limit for $h=0$.}\label{fig:mh-ferromagnetico}
\end{figure}

Let us now consider the curve for $m_\beta(h)=\medio{\am}_\beta(h)$ in the
thermodynamic limit. We compute it with the same method employed in Section~\ref{results-h},
but now we cannot apply the antisymmetrisation~\eqref{eq:odd}. The result is plotted
in Fig.~\ref{fig:mh-ferromagnetico}, where we plot $L=512$ for magnetic fields
in the range $h\in [0,10^{-3}]$. We also plot $L=256$ 
to check that both curves coincide and we have reached
the thermodynamic limit within our errors.
We appreciate in this figure just how efficient
the antisymmetrisation procedure was in Fig.~\ref{fig:mh}, which had 
much smaller errors.   

\begin{table}[h]
\centering
\caption{Canonical averages for several physical quantities
of an Ising lattice at $\beta=0.4473$ computed with 
the tethered method (T). The grid of 
$\hat m$ values is uniform  in the narrow simulated band. Also included
are the exact results for finite lattices from~\cite{FERDINAND-FISHER}
and the exact results in the thermodynamic limit from~\cite{ONSAGER,YANG}.
We appreciate that by simulating
only a very small range $\Delta \hat m$ of values for $\hat m$ we
can obtain very precise values. 
Within our error, we have already reached the thermodynamic limit for $L=512$.}
\label{tab:results-BSP}
\smallskip

\begin{tabular*}{\columnwidth}{@{\extracolsep{\fill}}rccllllll}
\hline
\multicolumn{1}{c}{\boldmath $L$} &
 \boldmath $N_\mathrm{points}$ & \boldmath $\Delta \hat m$ &
\multicolumn{1}{c}{\boldmath $-\medio{\au}_\beta$} & \multicolumn{1}{c}{\boldmath $C$}
&\multicolumn{1}{c}{\boldmath $\xi_2$}&\multicolumn{1}{c}{\boldmath $\medio{\am^2}_\beta$}
&\multicolumn{1}{c}{\boldmath $\medio{\am}_\beta$} &\multicolumn{1}{c}{\boldmath $\chi_2$}\\
\hline
128 (T) & 90 & 0.9  &  1.490\,397(18)        & 8.874(4) & 10.394(17) & 0.519\,87(8)& 0.719\,34(6) & 39.65(8)\\ 
128 (E) &    &      &  1.490\,409\,763\ldots & 8.877\,363\ldots\\
\hline
256 (T) & 79 & 0.39  &  1.490\,407(11)        & 8.869(5) & 11.26(4) & 0.518\,16(5) & 0.719\,41(4) & 39.36(7) \\
256 (E) &    &      &  1.490\,415\,672\ldots & 8.874\,075\ldots\\
\hline
512 (T) & 27 & 0.13 &  1.490\,419(5)         & 8.877(5)         &  11.5(3) & 0.517\,77(4) & 0.719\,45(3) & 39.37(9)\\
512 (E) &    &      &  1.490\,415\,689\ldots & 8.874\,046\ldots\\
\hline
1024 (T)& 27 & 0.13 &  1.490\,416(4)         & 8.868(7)         &  11.4(18)& 0.517\,64(4) & 0.719\,45(2) & 39.48(11)\\
1024 (E)&    &      &  1.490\,415\,689\ldots & 8.874\,046\ldots\\
\hline
$\infty$ (E) & &      &  1.490\,415\,689\ldots & 8.874\,046\ldots &          &              & 0.719\,436\ldots &        \\
\hline
\end{tabular*}   
\end{table}

\subsection{Ensemble equivalence in the ferromagnetic phase}\label{ensemble-equivalence}
Once we wander away from the critical point, the main objective
is finding the value of physical quantities in the thermodynamic limit.
The ensemble equivalence property suggests a way to reach this limit
without constructing the whole canonical $p(\hat m)$, but 
by concentrating instead on its maximum. From the computational 
point of view, this supposes a dramatic reduction in the needed
effort for a TMC simulation. 

Ensemble equivalence can be expressed in mathematical terms by 
\begin{equation}\label{eq:ensemble-equivalence}
\lim_{N\to\infty} \medio{\aO}_{\beta} = \lim_{N\to\infty} \medio{\aO}_{\medio{\hat m}_\beta, \beta}.
\end{equation}
This equation can be understood
as a more formal way of summarising the behaviour of Fig.~\ref{fig:Yang}. Indeed, 
we saw in the previous section that we could reconstruct the canonical 
averages considering only a very narrow range of $\hat m$; in the thermodynamic
limit a single point would be sufficient. 

For the Ising model we can situate this point exactly
because we know Yang's spontaneous magnetisation
\begin{equation}
\hat m_{\textsc{y}}(\beta) = \lim_{N \to \infty} \medio{\hatam}_{\beta} = 
\lim_{N \to \infty} \medio{\am}_{\beta} + \frac12 = \left[1- (\sinh 2\beta)^{-4}\right]^{1/8} + \frac12. 
\end{equation}
We could then run simulations for several lattice sizes at precisely $\hat m_\textsc{y}$ 
and study the evolution of $\medio{\aO}_{\hat m_\textsc{y},\beta}$ as we increase $L$.
This is not the most practical approach, as for a model other than the $D=2$ Ising
lattice we would not know the position of the peak beforehand. Instead, we will 
follow a more general analysis that would work in more complex situations. 

Let us consider the canonical average of some quantity 
and recall that we are using periodic boundary conditions, 
so the approach to the thermodynamic limit is exponential
\begin{equation}\label{eq:ey}
\medio{\aO}_\beta = \int_{1/2}^\infty \dd\hat m\ p(\hat m,\beta; L) \medio{\aO}_{\hat m} 
= O^\infty_\beta + A_{\aO} \ee^{-L/\xi_\infty},
\end{equation}
where $A_\aO$ is a constant amplitude. 
We have just considered the positive magnetisation peak.

Since the integral will be dominated by a saddle point at $\hat m_\text{peak}^+$,
with $\hat m_\text{peak}^+ \xrightarrow{L\to\infty} \hat m_\textsc{y}$,
we can approximate the pdf by a Gaussian
\begin{equation}
p(\hat m,\beta;L) \simeq \sqrt{\frac{N}{2\pi\chi_2}} \exp\left[-\frac{N(\hat m - \hat m_\text{peak})^2}{2\chi_2}\right].
\end{equation}
Therefore, we expect the tethered average of $\aO$
at this saddle point to approach the canonical average \eqref{eq:ey}, with a correction
of order $N^{-1}$,
\begin{equation}
O^\infty_\beta= \medio{\aO}_\beta - A_{\aO} \ee^{-L/\xi_\infty} = \medio{\aO}_{\hat m_\text{peak}^+, \beta} + \mathcal O\left(L^{-D}\right).
\end{equation}
To ease the notation we shall use the definition
\begin{equation}
O^+_{\text{peak},\beta} = \medio{\aO}_{\hat m_\text{peak}^+,\beta}.
\end{equation}
This simple analysis provides a practical way of approaching the thermodynamic
limit without knowing the limiting position of the peak in advance. 

We first run a complete simulation for some small lattice, covering the whole
range of $\hat m$. This provides a first approximation to the position
of the peak. For growing lattices, we just compute two or three points
at both sides of where we think the maximum is going to be. Our objective
is not to reconstruct the whole peak of $p(\hat m)$, just to 
find a good approximation to the point $\hat m_\text{peak}^+$ 
where $\medio{\hatah}_{\hat m,\beta}$ vanishes. We use the same
procedure as in Section~\ref{peaks}, finding the zero of the cubic
spline and interpolating the physical observables. Actually, if the position
of the peak is sufficiently bounded we could just place one point very closely 
at either side and use a linear interpolation.

With this procedure we are able to compute the tethered mean values 
of the relevant physical quantities at the peak with a minimum of
numerical effort. Here we shall apply this method to the energy
and we shall also characterise the approach of the peak to $\hat m_\textsc{y}$.
To the latter purpose, we have computed $h_\textsc{y}=\medio{\hatah}_{\hat m_y,\beta}$
for several lattice sizes and studied how fast it approaches zero.
We also give the values for the position of the peak (Table~\ref{tab:mYang}).

\begin{table}[t]
\centering
\caption{Tethered mean values of several parameters 
at the peak of the probability density function
for $\beta=0.4473$ and $\beta=0.6$, together with the 
value of $\hat h_\textsc{y}=\medio{\hatah}_{\hat m_\textsc{y}}$
(this observable is zero at the peak and helps characterise
how close we are to it). The exact value for 
an infinite lattice, which coincides with the canonical 
average, is also included for comparison.}\label{tab:mYang}
\smallskip

\begin{tabular*}{\columnwidth}{@{\extracolsep{\fill}}rcllccll}
\cline{2-8}
& \multicolumn{3}{c}{\boldmath $\beta = 0.4473$ } & &
 \multicolumn{3}{c}{\boldmath  $\beta = 0.6$} \\
\hline
\multicolumn{1}{c}{\boldmath $L$} &
 \boldmath $\hat h_\textsc{\bfseries Y}\cdot10^{5}$ & \multicolumn{1}{c}{\boldmath $\hat m_\mathrm{peak}^+$} &
                \multicolumn{1}{c}{\boldmath $- u_{\mathrm{peak},\beta}^+$} 
            &  & \boldmath $\hat h_\textsc{\bfseries Y}\cdot10^5$
              & \multicolumn{1}{c}{\boldmath $\hat m_\mathrm{peak}^+$} &
                \multicolumn{1}{c}{\boldmath $- u_{\mathrm{peak},\beta}^+$} \\
\hline
16  &$2984(3)$  & 1.343\,84(4) & 1.577\,07(8)  & &$ -290.6(14)$ & 1.471\,943(7)        & 1.912\,98(4)           \\ 
32  &$924.7(16)$& 1.299\,30(7) & 1.528\,98(9)  & &$ -52.5(8)  $ & 1.473\,299(5)        & 1.910\,18(2)                \\
64  & $284.9(9)$  & 1.263\,33(7) & 1.505\,48(5)  & &$ -11.0(4)  $ & 1.473\,543(2)    & 1.909\,374(9)   \\
128 & $86.0(6)$ & 1.239\,88(9)& 1.495\,63(4)  & &$ -2.45(19) $ & 1.473\,594\,0(12)& 1.909\,165(5)    \\
256 & $25.5(3)$  & 1.227\,32(8) & 1.491\,99(2) & &$ -0.67(11) $ & 1.473\,604\,7(2) & 1.909\,107(3)       \\
512 & $6.9(2)$  & 1.222\,04(6) & 1.490\,859(16)& &$ -0.19(5)  $ & 1.473\,607\,7(4) & 1.909\,090\,7(15)     \\
1024& $2.11(11)$  & 1.220\,24(4) & 1.490\,538(9) & &$ -0.03(2)$   & 1.473\,608\,3(2) & 1.909\,086\,7(4)\\
\hline
$\infty$ & 0    & 1.219\,435\ldots & 1.490\,416\ldots& &0  & 1.473\,608\,7\ldots  & 1.909\,086\,2\ldots\\
\hline
\end{tabular*}
\end{table}

Following the above analysis we should find that 
\begin{align}
|u^+_{\text{peak},\beta}-u^\infty_\beta | &= A_u\cdot L^{-\zeta_u},\nonumber\\
|\hat m^+_{\text{peak},\beta}-\hat m_\textsc{y} | &= A_{\hat m}\cdot L^{-\zeta_{\hat m}},\label{eq:zeta}\\
\hat h_\textsc{y}&= A_{\hat h}\cdot L^{-\zeta_{\hat h}}\nonumber,
\end{align}
with $\zeta\approx 2$. We have tried to fit these quantities 
to a power law for $\beta=0.4473$, but we found its behaviour 
to be more complex.  Instead, we present in Table~\ref{tab:ensemble-equivalence}
the result of taking the points for $L$ and $2L$ and computing
the effective exponent between them (that is, the value of
$\zeta$ so that the power law would pass exactly 
by those points). As we can see, our results 
are always $\zeta<2$, even though this exponent grows with $L$.

We believe this was caused by the proximity of the critical point, 
so we ran analogous simulations for $\beta=0.6$. We were able to complete
this new computations in very little time, following the above procedure.
For example, for $L=512$ the position of the peak was so well bounded
that we just computed one point at either side.

\begin{table}[t]
\centering
\caption{Rate at which several observables approach zero. We consider a functional
form $A\cdot L^{-\zeta}$ and compute the effective exponent $\zeta$
from the ratio of the computed values at consecutive lattice sizes. 
We consider three exponents, $\zeta_{\hat h}$, $\zeta_{\hat m}$ and $\zeta_u$
for the evolution of $\hat h_\textsc{y}$, $\hat m_\text{peak}^+$ and $u^+_\text{peak}$, 
respectively, see equation~\eqref{eq:zeta}. 
We observe that for $\beta=0.6$ the effective exponent approaches $2$, as
expected from the discussion in the text, while for $\beta=0.4473$ the 
proximity of the critical point complicates the analysis.
}\label{tab:ensemble-equivalence}
\smallskip

\begin{tabular*}{\columnwidth}{@{\extracolsep{\fill}}rlllclll}
\cline{2-8}
& \multicolumn{3}{c}{\boldmath $\beta = 0.4473$ } & & 
 \multicolumn{3}{c}{\boldmath  $\beta = 0.6$} \\
\hline
\multicolumn{1}{c}{\boldmath $L$} 
& \multicolumn{1}{c}{\boldmath $\zeta_{\hat h}$}
              & \multicolumn{1}{c}{\boldmath $\zeta_{\hat m}$}
              & \multicolumn{1}{c}{\boldmath $\zeta_u$} & 
              & \multicolumn{1}{c}{\boldmath $\zeta_{\hat h}$}
              & \multicolumn{1}{c}{\boldmath $\zeta_{\hat m}$} &
                \multicolumn{1}{c}{\boldmath $\zeta_u$} \\
\hline
16   & 1.690(3) & 0.6394(14)& 1.168(4) & & 2.47(2)  & 2.43(2)  & 1.83(3)\\
32   & 1.699(5)& 0.864(3)  & 1.356(6) & & 2.25(5)  & 2.24(5)  & 1.93(5)\\
64   & 1.729(12)  & 1.102(7)  & 1.531(12)& & 2.17(12) & 2.16(13) & 1.87(10)\\
128  & 1.75(2)  & 1.375(16) & 1.73(3)  & & 1.9(3)   & 1.89(13) & 1.9(2)  \\
256  & 1.88(5)  & 1.60(4)   & 1.83(6)  & & 1.8(4)   & 2.0(5)   & 2.2(5) \\
512  & 1.71(9) & 1.70(8)   & 1.85(12) & & 2.7(12)  & 1.4(10)  & 3.1(12)\\
\hline           
\end{tabular*}
\smallskip
\end{table}

Comparing Table~\ref{tab:mYang}
with Table~\ref{tab:results-Tc} we see that for $\beta=0.6$, with a computation effort 
almost $40$ times smaller, we have obtained a result
an order of magnitude more precise than what we had at $\beta_{\mathrm{c}}$. 
Recomputing the effective exponents for these new simulations we obtain 
results compatible with $\zeta=2$. Notice that for this temperature
the error in the exponents is much bigger than that for $\beta=0.4473$.
The reason is clear from Table~\ref{tab:mYang}. The left hand
sides of equations~\eqref{eq:zeta} are now much closer to zero
than in $\beta=0.4473$, yet their errors are only slightly smaller.
In the computation of the effective exponents only  the relative
errors matter, which explains our bigger uncertainties. Notice, however,
that we have been able to distinguish values for $h_\textsc{y}$ of 
order $10^{-6}$ from zero and that we have located the peak
with seven significant figures.

\section{Conclusions and outlook}\label{conclusions}
We have presented the Tethered Monte Carlo method, a 
completely unspecific formalism to reconstruct
the effective potential of the order parameter.
This method is based on a new
statistical ensemble, which we have described 
in detail. The tethered ensemble not only 
allows to reproduce the canonical results, but
is also specially suited to the study 
of the broken symmetry phase and 
the effects of an external magnetic field.   

We have implemented this formalism in the $D=2$ Ising model, were we 
can make exhaustive  checks of our results, either against 
exact solutions or high precision computations performed 
with canonical cluster algorithms.
We have not have tried to optimise the efficiency for this 
particular model, choosing instead the most generally
appliable method: a standard local Metropolis algorithm. 
The possibility of developing a cluster algorithm for the tethered formalism
will be considered in the future.

The method has been tested to a high degree of accuracy
in a large variety of situations:
the critical point, the scaling paramagnetic region,
the thermodynamic limit in the ferromagnetic region
and the nonlinear response to an external magnetic fields.
Our Monte Carlo implementation has proven to be
remarkably efficient for the ferromagnetic phase
and specially for computations with an external
magnetic field.

One of its most conspicuous features is the 
absence of critical slowing down for all
functions of the magnetisation, even though
we have used a local algorithm. This is 
probably due to the fact that the tethered ensemble
instantaneously propagates information
to the whole lattice by means of 
a `magnetic bath'. Our algorithm allows 
us to study each magnetisation independently
without having to wander randomly in the
magnetisation space, as is the case
for multicanonical or Wang-Landau computations.

We expect Tethered Monte Carlo to be of
great help whenever large tunnelling 
barriers appear, associated to
observables not present in the Hamiltonian. A non exhaustive
list of instances are the standard magnetisation
in the Random Field Ising Model~\cite{NATTERMAN}, the staggered
magnetisation for the Diluted Antiferromagnet in a Field~\cite{DAFF} or
the Polyakov loop for lattice gauge theories~\cite{GAUGE}. 
In these cases `tethering' extensive quantities
other than the magnetisation could prove
an invaluable guide for the exploration of phase space.
In particular, we wish to apply this formalism in the near future
to undertake a thorough study of the condensation 
transition~\cite{JANKE-CONDENSACION,CONDENSACION}.

\section*{Acknowledgments}
We were partially supported by MEC (Spain) through contract No. FIS-2006-08533-C03-01
and by CAM (Spain) through contract No. CCG07-UCM/ESP-2532.

\appendix
\section{Some numerical details}\label{detalles-tecnicos}
A numerical implementation of equation \eqref{eq:canonical-average} can be 
done in several different ways of equivalent numerical accuracy. Here we 
briefly explain our choices. 

Recall that we have represented $\medio{\hatah}_{\hat m,\beta}$ 
with a cubic spline interpolation. We have not used the so called
natural spline, which imposes vanishing curvature for $\medio{\hatah}_{\hat m,\beta}$
at $\hat m_\text{max}$ and $\hat m_\text{min}$. Instead, we 
have estimated the derivative of this function at both ending points
with a parabolic interpolation. To compute the canonical average
of~\eqref{eq:canonical-average} we also represent $\medio{\aO}_{\hat m,\beta}$
with a cubic spline. However, naively applying this interpolation scheme
for the exponential, $p(\hat m)=\exp[ N \varOmega_N(\hat m)]$,
could introduce strong integration errors. Fortunately, 
this can be easily solved by accurately representing the exponent $N\varOmega_N$, which
is a smooth function (recall the curve in logarithmic scale of Fig.~\ref{fig:tethered}).

$\varOmega_N(\hat m,\beta)$, being the integral of the cubic spline for $\medio{\hatah}_{\hat m, \beta}$,
is a fourth order polynomial between each pair of consecutive points in the 
$\hat m$ grid, which can be exactly computed from the coefficients
of the cubic spline. To avoid losing precision, 
we evaluate $\varOmega_N(\hat m,\beta)$ at an extended grid
that includes $3$ equally spaced intermediate points between each 
pair of simulated values of $\hat m$.  This way the Lagrange 
interpolating polynomial for each segment of 
the extended lattice (two original points plus three
intermediate ones) represents the exact integral of our spline.

Of course, the pdf,
$\exp[N \varOmega_N(\hat m, \beta)]$, is not a polynomial anymore. It is, 
however, a smooth function so a self consistent 
Romberg method~\cite{NR} provides an estimate of the integral~\eqref{eq:canonical-average}
with any required numerical accuracy. Notice that this yields
the basically  exact results for a given interpolation 
of $\medio{\hatah}_{\hat m,\beta}$, but it does not cure
any discretisation errors introduced by the spline.

Typically, even with a very moderate effort, the Romberg integration error
has been much smaller than the statistical one. There is one exception:
the fluctuation-dissipation formulas, such as~\eqref{eq:calor-especifico}, because of 
the large cancellations between the two terms.
To solve this problem, we have computed the fluctuation-dissipation formula 
as a sum of two squares:
\begin{align}
N^{-1} C= \medio{\au^2}_\beta - \medio{\au}_\beta^2 &= 
      \int\dd \hat m\ p(\hat m) \left[ \medio{\au^2}_{\hat m} - \medio{\au}^2_{\hat m}
                                 + \medio{\au}^2_{\hat m} - \medio{\au}^2_\beta\right]\nonumber\\
         &= \int\dd\hat m\ p(\hat m) \left[ \medio{( \au -\medio{\au}_{\hat m})^2}_{\hat m}
                                + \left(\medio{\au}_{\hat m} - \medio{\au}_\beta\right)^2                           \right].
\end{align}
In spite of this, as a consistency check, we have also employed the original equation 
and forced the Romberg integral to yield the same value, by 
reducing its tolerance.

\end{document}